\documentclass[nofootinbib,reprint,superscriptaddress]{revtex4-2}
\usepackage{amsmath}
\usepackage{amsfonts}
\usepackage{amssymb}
\usepackage{bm}
\usepackage{diagbox}
\usepackage{enumitem}
\usepackage{graphicx}
\usepackage[colorlinks,citecolor=blue,filecolor=blue,linkcolor=blue,urlcolor=blue]{hyperref}
\usepackage[all]{hypcap}
\usepackage[utf8]{inputenc}
\usepackage{lineno}
\usepackage[caption=false]{subfig}
\usepackage{threeparttable}

\begin{document}
\allowdisplaybreaks

\title{Statistical uncertainty estimation of higher-order cumulants with finite efficiency and its application in heavy-ion collisions}
\author{Fan Si}\affiliation{State Key Laboratory of Particle Detection and Electronics, University of Science and Technology of China, Hefei 230026, China}
\author{Yifei Zhang}\email{ephy@ustc.edu.cn}\affiliation{State Key Laboratory of Particle Detection and Electronics, University of Science and Technology of China, Hefei 230026, China}

\begin{abstract}

We derive the general analytical expressions for the statistical uncertainties of cumulants up to fourth order including an efficiency correction. The analytical expressions have been tested with a toy Monte Carlo model analysis. An application to the study of particle multiplicity fluctuations in heavy-ion collisions is investigated. In this derivation, a mathematical proof is given that the validity of an averaged efficiency correction and the fluctuations induced by the non-uniformity of efficiency can be eliminated. The estimation of statistical uncertainties using the analytical formulas is found to be significantly faster than the commonly used bootstrap method. The simplicity and
efficiency of using the analytical formulas may be useful for massive data analysis in many fields.

\end{abstract}

\maketitle

\section{Introduction}

The cumulants of a probability distribution and their calculation have been thoroughly studied in probability theory and statistics~\cite{Kendall:1994ats}. The first two cumulants are the mean ($\mu$) and variance ($\sigma^2$). The third one is the same as the third central moment ($S\sigma^3$, $S$ for skewness), but the fourth ($\kappa\sigma^4$, $\kappa$ for kurtosis) and higher-order cumulants are not equal to the central moments. Cumulants are a series of additive statistics: for several independent stochastic variables, the cumulant of their sum is equal to the sum of their individual cumulants. Since the cumulants higher than second order of the Gaussian distribution remain zero, they are useful for the description of non-Gaussian fluctuations. Due to their unique properties, cumulants have been widely used in many fields, such as the fluctuation research in heavy-ion physics~\cite{Luo:2017faz,Luo:2020pef}, wave spectrum and signal analysis in electronic technology~\cite{Akbarizadeh:2012ans,Wang:2012msl,Peng:2019mcb}, biomedical signal processing~\cite{Acharya:2011ade,Noronha:2014acg}, and the application of economic statistics~\cite{Martin:2013cba,Pender:2015tnd,Zhang:2020mer}. The statistical uncertainty estimation of cumulants has been studied previously in Ref.~\cite{Kendall:1994ats}. However, due to the complexity of the analytical derivations, as an alternative, the statistical uncertainties are commonly estimated via a bootstrap method~\cite{Efron:1979cts} through computationally intensive Monte Carlo resampling, and its calculation time cost is proportional to the scale of the data sample and the number of resamplings.

In particular, in many cases of data collection and measurements of statistics, the objects of interests are observed in imperfection with a probability less than unity, namely, measuring efficiency, and the recorded distribution is distorted compared with the original one as a result. This is called the finite-efficiency effect. For example, the limited capability of the detector system in heavy-ion collision experiments results in lost tracks and missing particles with the non-uniform efficiency depending on many factors, such as the collision centrality and particle kinematic parameters (transverse momentum, azimuth angle, and rapidity)~\cite{Anderson:2003ur}, which strongly affect the measured event-by-event particle multiplicity distribution. The measured statistics describing higher-order fluctuations and their statistical uncertainties are more sensitive to the efficiency, so an efficiency correction~\cite{Nonaka:2017kko,Kitazawa:2017ljq,Luo:2018ofd} should be carefully applied to estimate the true values of higher-order cumulants of the original distribution. However, since the statistical uncertainty estimation of the efficiency-corrected higher-order cumulants becomes much more complicated, no analytical formula derivation has been given to solve this issue so far. Although the statistical uncertainties can be estimated via the bootstrap method~\cite{Luo:2018ofd} costing lots of CPU time, their components and how they are essentially affected are incomprehensible. As an example, the final-state conserved quantities in heavy-ion collisions, like net-charge, net-strangeness, and net-baryon multiplicities, have been proposed as signatures for the phase-transition critical point between the quark-gluon plasma (QGP) phase and hadronic gas in the quantum chromodynamics (QCD) phase diagram~\cite{STAR:2014egu,STAR:2017tfy,STAR:2020tga}. A first glance at the non-monotonic distributions of the net-proton multiplicities as a function of the collision energy from the RHIC BES-I program has been given by the STAR Collaboration with large uncertainties, in particular in the high-baryon-density region near the possible critical area~\cite{STAR:2020tga,STAR:2021iop}. Thus, the estimation of statistical uncertainties with the imperfect and non-uniform efficiency correction is crucial to the significance of this observation.

In a previous study~\cite{Si:2021mdj}, the validity of an averaged efficiency correction for higher-order cumulants was analytically proved, and in a toy Monte Carlo model analysis, the statistical fluctuations of the efficiency-corrected cumulants were observed to depend on the non-uniformity of the valid efficiency employed in the efficiency correction, namely, correcting efficiency. In this paper, the analytical formulas for the statistical uncertainty estimation of efficiency-corrected higher-order cumulants are derived and then mathematically decomposed into their principle components to study their functional form and dependence on critical parameters such as the efficiency. A toy Monte Carlo model analysis is employed to check the analytical results, and a comparison between the analytical method and the bootstrap method is given.

\section{Assumption and definition}

\subsection{Binomial efficiency assumption}

In this paper, the finite measuring efficiency is assumed to be binomial efficiency, and, accordingly, the efficiency correction is employed with the binomial model. For example, there exist objects of interest where the number $X$ is a non-negative integer stochastic variable following the probability distribution function $P(X)$. The binomial efficiency $\alpha\in[0,1]$ makes that each individual object is observed independently with probability $\alpha$, and, thus, the measured number $x$ of objects follows
\begin{equation}
\label{eqb1}
\tilde{P}(x)=\sum_{X}P(X)\mathcal{B}_{X,\alpha}(x),
\end{equation}
where $\mathcal{B}_{X,\alpha}(x)$ denotes the binomial distribution defined by
\begin{equation}
\mathcal{B}_{X,\alpha}(x)=\frac{X!}{x!(X-x)!}\alpha^{x}(1-\alpha)^{X-x}
\end{equation}
with $\displaystyle{\alpha=\frac{\langle x\rangle}{\langle X\rangle}}$.

In heavy-ion collisions, the non-binomial effects are estimated by the efficiency correction with the unfolding method rather than the binomial model, and results shows that even if the maximum non-binomial effects are expected, their contribution can be quoted by the binomial efficiency correction within uncertainties~\cite{STAR:2021iop}.

\subsection{Phase-space definition}

In this paper, the definition of a phase space is the same as Ref.~\cite{Si:2021mdj}. Suppose that there exist $M$ series of objects of interest, whose numbers $\bm{X}=\left(X_1,X_2,\cdots,X_M\right)$ are non-negative integer stochastic variables following the probability distribution function $P(\bm{X})$, and their total number $\displaystyle{X=\sum_{i=1}^M X_i}$ follows $P'(X)$. A phase space is defined to contain these series of objects if
\begin{equation}
\label{eqb2}
P(\bm{X})=\sum_{X}P'(X)\mathcal{M}_{X,\bm{p}}(\bm{X}),
\end{equation}
where $\mathcal{M}_{X,\bm{p}}(\bm{X})$ denotes the multinomial distribution defined as
\begin{equation}
\mathcal{M}_{X,\bm{p}}(\bm{X})=\frac{X!}{\displaystyle{\prod_{i=1}^M X_i!}}\prod_{i=1}^M p_i^{X_i}
\end{equation}
with $\bm{p}=\left(p_1,p_2,\cdots,p_M\right)$ and $\displaystyle{p_i=\frac{\left\langle X_i\right\rangle}{\langle X\rangle}}$, in other words, objects in a phase space are produced with the total number $X$ determined by a certain distribution $P(X)$ and then allocated into each series by a certain probability vector $\bm{p}$.
For example, if there are independent $X_i\sim\mathrm{Poisson}\left(\lambda_i\right)$, Eq.~\eqref{eqb2} holds for $\displaystyle{X\sim\mathrm{Poisson}\left(\sum\lambda_i\right)}$.

In heavy-ion collisions, the phase space under this definition needs to be further studied.

\section{Covariances of estimated cumulants}

The moments and cumulants for a bivariant probability distribution function $P(X,Y)$, for example, are defined by their generating functions as~\cite{Kendall:1994ats}
\begin{align}
\left\langle X^kY^l\right\rangle&=\left.\partial_\theta^k\partial_\eta^lG(\theta,\eta)\right|_{\theta=\eta=0},\\
G(\theta,\eta)&=\sum_{X,Y}{P(X,Y)\mathrm{e}^{\theta X+\eta Y}}=\left\langle\mathrm{e}^{\theta X+\eta Y}\right\rangle,\\
\left\langle X^kY^l\right\rangle_\mathrm{c}&=\left.\partial_\theta^k\partial_\eta^lG_\mathrm{c}(\theta,\eta)\right|_{\theta=\eta=0},\\
G_\mathrm{c}(\theta,\eta)&=\ln{G(\theta,\eta)}=\ln{\left\langle\mathrm{e}^{\theta X+\eta Y}\right\rangle},
\end{align}
respectively, where $\partial_\theta$ represents $\partial/\partial\theta$.

The relation between cumulants and moments can be obtained from
\begin{equation}
\begin{split}
\left\langle X^kY^l\right\rangle_\mathrm{c}&=\left.\partial_\theta^k\partial_\eta^lG_\mathrm{c}(\theta,\eta)\right|_{\theta=\eta=0}\\
&=\left.\partial_\theta^k\partial_\eta^l\ln{G(\theta,\eta)}\right|_{\theta=\eta=0}.
\end{split}
\end{equation}
For example,
\begin{equation}
\begin{split}
\langle XY\rangle_\mathrm{c}&=\left.\partial_\theta\partial_\eta G_\mathrm{c}(\theta,\eta)\right|_{\theta=\eta=0}\\
&=\left.\left(\frac{\partial_\theta\partial_\eta G}{G}-\frac{\left(\partial_\theta G\right)\left(\partial_\eta G\right)}{G^2}\right)\right|_{\theta=\eta=0}\\
&=\langle XY\rangle-\langle X\rangle\langle Y\rangle.
\end{split}
\end{equation}

For moment estimation, the sample moments are used to estimate the population moments, and then the covariance of two estimated moments is obtained with~\cite{Kendall:1994ats}
\begin{widetext}
\begin{equation}
\label{eqa1}
n\cdot\mathrm{Cov}\left(\widehat{\left\langle X_1X_2\cdots X_k\right\rangle},\widehat{\left\langle Y_1Y_2\cdots Y_l\right\rangle}\right)=\left\langle X_1X_2\cdots X_kY_1Y_2\cdots Y_l\right\rangle-\left\langle X_1X_2\cdots X_k\right\rangle\left\langle Y_1Y_2\cdots Y_l\right\rangle,
\end{equation}
where $X_i$ ($i$ = 1, 2, $\cdots$, $k$) and $Y_j$ ($j$ = 1, 2, $\cdots$, $l$) are stochastic variables and $n$ denotes the number of samples. The hat ($\widehat{\cdot}$) represents an estimated quantity obtained from samples.

With the relation between cumulants and moments and Eq.~\eqref{eqa1}, the covariances (Cov) and variances (Var) of estimated cumulants can be obtained~\cite{Kendall:1994ats}. The formulas for cumulants up to fourth order are given below ($P$, $Q$, $R$, $S$, $T$ and $U$ are stochastic variables with large $n$):

\begin{align}
\text{First-order cumulants:}&\nonumber\\
n\cdot\mathrm{Cov}\left(\widehat{\left\langle P\right\rangle_\mathrm{c}},\widehat{\left\langle Q\right\rangle_\mathrm{c}}\right)&=n\cdot\mathrm{Cov}\left(\widehat{\left\langle P\right\rangle},\widehat{\left\langle Q\right\rangle}\right)=\left\langle PQ\right\rangle-\left\langle P\right\rangle\left\langle Q\right\rangle=\left\langle PQ\right\rangle_\mathrm{c},\\
n\cdot\mathrm{Cov}\left(\widehat{\left\langle P^2\right\rangle_\mathrm{c}},\widehat{\left\langle Q\right\rangle_\mathrm{c}}\right)&=\left\langle P^2Q\right\rangle_\mathrm{c},\\
n\cdot\mathrm{Cov}\left(\widehat{\left\langle PQ\right\rangle_\mathrm{c}},\widehat{\left\langle R\right\rangle_\mathrm{c}}\right)&=\left\langle PQR\right\rangle_\mathrm{c},\\
n\cdot\mathrm{Cov}\left(\widehat{\left\langle P^3\right\rangle_\mathrm{c}},\widehat{\left\langle Q\right\rangle_\mathrm{c}}\right)&=\left\langle P^3Q\right\rangle_\mathrm{c},\\
n\cdot\mathrm{Cov}\left(\widehat{\left\langle PQR\right\rangle_\mathrm{c}},\widehat{\left\langle S\right\rangle_\mathrm{c}}\right)&=\left\langle PQRS\right\rangle_\mathrm{c},\\
n\cdot\mathrm{Cov}\left(\widehat{\left\langle P^4\right\rangle_\mathrm{c}},\widehat{\left\langle Q\right\rangle_\mathrm{c}}\right)&=\left\langle P^4Q\right\rangle_\mathrm{c},\\
n\cdot\mathrm{Cov}\left(\widehat{\left\langle PQRS\right\rangle_\mathrm{c}},\widehat{\left\langle T\right\rangle_\mathrm{c}}\right)&=\left\langle PQRST\right\rangle_\mathrm{c}.\\
\text{Second-order cumulants:}&\nonumber\\
n\cdot\mathrm{Cov}\left(\widehat{\left\langle P^2\right\rangle_\mathrm{c}},\widehat{\left\langle Q^2\right\rangle_\mathrm{c}}\right)&=\left\langle P^2Q^2\right\rangle_\mathrm{c}+2\left\langle PQ\right\rangle_\mathrm{c}^2,\\
n\cdot\mathrm{Cov}\left(\widehat{\left\langle PQ\right\rangle_\mathrm{c}},\widehat{\left\langle RS\right\rangle_\mathrm{c}}\right)&=\left\langle PQRS\right\rangle_\mathrm{c}+\left\langle PR\right\rangle_\mathrm{c}\left\langle QS\right\rangle_\mathrm{c}+\left\langle PS\right\rangle_\mathrm{c}\left\langle QR\right\rangle_\mathrm{c},\\
n\cdot\mathrm{Cov}\left(\widehat{\left\langle P^3\right\rangle_\mathrm{c}},\widehat{\left\langle Q^2\right\rangle_\mathrm{c}}\right)&=\left\langle P^3Q^2\right\rangle_\mathrm{c}+6\left\langle P^2Q\right\rangle_\mathrm{c}\left\langle PQ\right\rangle_\mathrm{c},\\
\begin{split}
n\cdot\mathrm{Cov}\left(\widehat{\left\langle PQR\right\rangle_\mathrm{c}},\widehat{\left\langle ST\right\rangle_\mathrm{c}}\right)&=\left\langle PQRST\right\rangle_\mathrm{c}+\left\langle PQS\right\rangle_\mathrm{c}\left\langle RT\right\rangle_\mathrm{c}+\left\langle PQT\right\rangle_\mathrm{c}\left\langle RS\right\rangle_\mathrm{c}+\left\langle PRS\right\rangle_\mathrm{c}\left\langle QT\right\rangle_\mathrm{c}\\
&\hphantom{\quad\ }+\left\langle PRT\right\rangle_\mathrm{c}\left\langle QS\right\rangle_\mathrm{c}+\left\langle QRS\right\rangle_\mathrm{c}\left\langle PT\right\rangle_\mathrm{c}+\left\langle QRT\right\rangle_\mathrm{c}\left\langle PS\right\rangle_\mathrm{c},
\end{split}\\
n\cdot\mathrm{Cov}\left(\widehat{\left\langle P^4\right\rangle_\mathrm{c}},\widehat{\left\langle Q^2\right\rangle_\mathrm{c}}\right)&=\left\langle P^4Q^2\right\rangle_\mathrm{c}+8\left\langle P^3Q\right\rangle_\mathrm{c}\left\langle PQ\right\rangle_\mathrm{c}+6\left\langle P^2Q\right\rangle_\mathrm{c}^2,\\
\begin{split}
n\cdot\mathrm{Cov}\left(\widehat{\left\langle PQRS\right\rangle_\mathrm{c}},\widehat{\left\langle TU\right\rangle_\mathrm{c}}\right)&= \left\langle PQRSTU\right\rangle_\mathrm{c}+\left\langle PQRT\right\rangle_\mathrm{c}\left\langle SU\right\rangle_\mathrm{c}+\left\langle PQST\right\rangle_\mathrm{c}\left\langle RU\right\rangle_\mathrm{c}+\left\langle PRST\right\rangle_\mathrm{c}\left\langle QU\right\rangle_\mathrm{c}\\
&\hphantom{\quad\ }+\left\langle QRST\right\rangle_\mathrm{c}\left\langle PU\right\rangle_\mathrm{c}+\left\langle PQRU\right\rangle_\mathrm{c}\left\langle ST\right\rangle_\mathrm{c}+\left\langle PQSU\right\rangle_\mathrm{c}\left\langle RT\right\rangle_\mathrm{c}+\left\langle PRSU\right\rangle_\mathrm{c}\left\langle QT\right\rangle_\mathrm{c}\\
&\hphantom{\quad\ }+\left\langle QRSU\right\rangle_\mathrm{c}\left\langle PT\right\rangle_\mathrm{c}+\left\langle PQT\right\rangle_\mathrm{c}\left\langle RSU\right\rangle_\mathrm{c}+\left\langle PRT\right\rangle_\mathrm{c}\left\langle QSU\right\rangle_\mathrm{c}+\left\langle PST\right\rangle_\mathrm{c}\left\langle QRU\right\rangle_\mathrm{c}\\
&\hphantom{\quad\ }+\left\langle PQU\right\rangle_\mathrm{c}\left\langle RST\right\rangle_\mathrm{c}+\left\langle PRU\right\rangle_\mathrm{c}\left\langle QST\right\rangle_\mathrm{c}+\left\langle PSU\right\rangle_\mathrm{c}\left\langle QRT\right\rangle_\mathrm{c}.
\end{split}\\
\text{Third-order cumulants:}&\nonumber\\
n\cdot\mathrm{Cov}\left(\widehat{\left\langle P^3\right\rangle_\mathrm{c}},\widehat{\left\langle Q^3\right\rangle_\mathrm{c}}\right)&=\left\langle P^3Q^3\right\rangle_\mathrm{c}+9\left\langle P^2Q^2\right\rangle_\mathrm{c}\left\langle PQ\right\rangle_\mathrm{c}+9\left\langle PQ^2\right\rangle_\mathrm{c}\left\langle P^2Q\right\rangle_\mathrm{c}+6\left\langle PQ\right\rangle_\mathrm{c}^3,\\
\begin{split}
n\cdot\mathrm{Cov}\left(\widehat{\left\langle PQR\right\rangle_\mathrm{c}},\widehat{\left\langle STU\right\rangle_\mathrm{c}}\right)&=\left\langle PQRSTU\right\rangle_\mathrm{c}+\left\langle PQST\right\rangle_\mathrm{c}\left\langle RU\right\rangle_\mathrm{c}+\left\langle PQSU\right\rangle_\mathrm{c}\left\langle RT\right\rangle_\mathrm{c}+\left\langle PQTU\right\rangle_\mathrm{c}\left\langle RS\right\rangle_\mathrm{c}\\
&\hphantom{\quad\ }+\left\langle PRST\right\rangle_\mathrm{c}\left\langle QU\right\rangle_\mathrm{c}+\left\langle PRSU\right\rangle_\mathrm{c}\left\langle QT\right\rangle_\mathrm{c}+\left\langle PRTU\right\rangle_\mathrm{c}\left\langle QS\right\rangle_\mathrm{c}+\left\langle QRST\right\rangle_\mathrm{c}\left\langle PU\right\rangle_\mathrm{c}\\
&\hphantom{\quad\ }+\left\langle QRSU\right\rangle_\mathrm{c}\left\langle PT\right\rangle_\mathrm{c}+\left\langle QRTU\right\rangle_\mathrm{c}\left\langle PS\right\rangle_\mathrm{c}+\left\langle PQS\right\rangle_\mathrm{c}\left\langle RTU\right\rangle_\mathrm{c}+\left\langle PQT\right\rangle_\mathrm{c}\left\langle RSU\right\rangle_\mathrm{c}\\
&\hphantom{\quad\ }+\left\langle PQU\right\rangle_\mathrm{c}\left\langle RST\right\rangle_\mathrm{c}+\left\langle PRS\right\rangle_\mathrm{c}\left\langle QTU\right\rangle_\mathrm{c}+\left\langle PRT\right\rangle_\mathrm{c}\left\langle QSU\right\rangle_\mathrm{c}+\left\langle PRU\right\rangle_\mathrm{c}\left\langle QST\right\rangle_\mathrm{c}\\
&\hphantom{\quad\ }+\left\langle QRS\right\rangle_\mathrm{c}\left\langle PTU\right\rangle_\mathrm{c}+\left\langle QRT\right\rangle_\mathrm{c}\left\langle PSU\right\rangle_\mathrm{c}+\left\langle QRU\right\rangle_\mathrm{c}\left\langle PST\right\rangle_\mathrm{c}+\left\langle PS\right\rangle_\mathrm{c}\left\langle QT\right\rangle_\mathrm{c}\left\langle RU\right\rangle_\mathrm{c}\\
&\hphantom{\quad\ }+\left\langle PS\right\rangle_\mathrm{c}\left\langle QU\right\rangle_\mathrm{c}\left\langle RT\right\rangle_\mathrm{c}+\left\langle PT\right\rangle_\mathrm{c}\left\langle QS\right\rangle_\mathrm{c}\left\langle RU\right\rangle_\mathrm{c}+\left\langle PT\right\rangle_\mathrm{c}\left\langle QU\right\rangle_\mathrm{c}\left\langle RS\right\rangle_\mathrm{c}\\
&\hphantom{\quad\ }+\left\langle PU\right\rangle_\mathrm{c}\left\langle QS\right\rangle_\mathrm{c}\left\langle RT\right\rangle_\mathrm{c}+\left\langle PU\right\rangle_\mathrm{c}\left\langle QT\right\rangle_\mathrm{c}\left\langle RS\right\rangle_\mathrm{c},
\end{split}\\
\begin{split}
n\cdot\mathrm{Cov}\left(\widehat{\left\langle P^4\right\rangle_\mathrm{c}},\widehat{\left\langle Q^3\right\rangle_\mathrm{c}}\right)&=\left\langle P^4Q^3\right\rangle_\mathrm{c}+12\left\langle P^3Q^2\right\rangle_\mathrm{c}\left\langle PQ\right\rangle_\mathrm{c}+12\left\langle P^3Q\right\rangle_\mathrm{c}\left\langle PQ^2\right\rangle_\mathrm{c}\\
&\hphantom{\quad\ }+18\left\langle P^2Q^2\right\rangle_\mathrm{c}\left\langle P^2Q\right\rangle_\mathrm{c}+36\left\langle P^2Q\right\rangle_\mathrm{c}\left\langle PQ\right\rangle_\mathrm{c}^2,
\end{split}\\
\begin{split}
n\cdot\mathrm{Cov}\left(\widehat{\left\langle P^4\right\rangle_\mathrm{c}},\widehat{\left\langle P^2Q\right\rangle_\mathrm{c}}\right)&=\left\langle P^6Q\right\rangle_\mathrm{c}+4\left\langle P^5\right\rangle_\mathrm{c}\left\langle PQ\right\rangle_\mathrm{c}+8\left\langle P^4Q\right\rangle_\mathrm{c}\left\langle P^2\right\rangle_\mathrm{c}+14\left\langle P^4\right\rangle_\mathrm{c}\left\langle P^2Q\right\rangle_\mathrm{c}\\
&\hphantom{\quad\ }+16\left\langle P^3Q\right\rangle_\mathrm{c}\left\langle P^3\right\rangle_\mathrm{c}+24\left\langle P^3\right\rangle_\mathrm{c}\left\langle P^2\right\rangle_\mathrm{c}\left\langle PQ\right\rangle_\mathrm{c}+12\left\langle P^2Q\right\rangle_\mathrm{c}\left\langle P^2\right\rangle_\mathrm{c}^2.
\end{split}\\
\text{Fourth-order cumulants:}&\nonumber\\
\begin{split}
n\cdot\mathrm{Var}\left(\widehat{\left\langle P^4\right\rangle_\mathrm{c}}\right)&=\left\langle P^8\right\rangle_\mathrm{c}+16\left\langle P^6\right\rangle_\mathrm{c}\left\langle P^2\right\rangle_\mathrm{c}+48\left\langle P^5\right\rangle_\mathrm{c}\left\langle P^3\right\rangle_\mathrm{c}+34\left\langle P^4\right\rangle_\mathrm{c}^2\\
&\hphantom{\quad\ }+72\left\langle P^4\right\rangle_\mathrm{c}\left\langle P^2\right\rangle_\mathrm{c}^2+144\left\langle P^3\right\rangle_\mathrm{c}^2\left\langle P^2\right\rangle_\mathrm{c}+24\left\langle P^2\right\rangle_\mathrm{c}^4.
\end{split}
\end{align}

Here the covariances of estimated cumulants in the univariate case are summarized as
\begin{equation}
n\cdot\mathrm{Cov}\left(\widehat{\left\langle X^k\right\rangle_\mathrm{c}},\widehat{\left\langle X^l\right\rangle_\mathrm{c}}\right)=\sum_{r=1}^{\min(k,l)}{f_r(k,l)},
\end{equation}
where $f_r(k,l)$ denotes a product of $r$ cumulants defined by
\begin{align}
f_1(k,l)&=\left\langle X^{k+l}\right\rangle_\mathrm{c},\\
f_r(k,l)&=\sum_{\tiny\substack{i_1,i_2,\cdots,i_r\geq2\\i_1+i_2+\cdots+i_r=k+l}}{\frac{g_r\left(k,l;i_1,i_2,\cdots,i_r\right)}{\Delta_{i_1,i_2,\cdots,i_r}}}\left\langle X^{i_1}\right\rangle_\mathrm{c}\left\langle X^{i_2}\right\rangle_\mathrm{c}\cdots\left\langle X^{i_r}\right\rangle_\mathrm{c},
\end{align}
with the numerical coefficients
\begin{equation}
g_r\left(k,l;i_1,i_2,\cdots,i_r\right)=\sum_{\tiny\substack{1\leq j_1\leq i_1-1\\\cdots\\1\leq j_r\leq i_r-1\\j_1+j_2+\cdots+j_r=k}}\left[\left(\substack{k\\j_1}\right)\left(\substack{l\\i_1-j_1}\right)\right]\left[\left(\substack{k-j_1\\j_2}\right)\left(\substack{l-i_1+j_1\\i_2-j_2}\right)\right]\cdots\left[\left(\substack{j_r\\j_r}\right)\left(\substack{i_r-j_r\\i_r-j_r}\right)\right],
\end{equation}
and $\Delta_{i_1,i_2,\cdots,i_r}$ denotes the factorial product of counts of the same numbers in $i_1,i_2,\cdots,i_r$, for example, $\Delta_{2,2,2,3}=3!\times1!$ and $\Delta_{2,2,3,3}=2!\times2!$.

In the multivariate case,
\begin{equation}
n\cdot\mathrm{Cov}\left(\widehat{\left\langle X_1X_2\cdots X_k\right\rangle_\mathrm{c}},\widehat{\left\langle Y_1Y_2\cdots Y_l\right\rangle_\mathrm{c}}\right)=\sum_{r=1}^{\min(k,l)}{f'_r(k,l)},
\end{equation}
where $f'_r(k,l)$ denotes a product of $r$ cumulants meeting the following requirements:
\begin{itemize}[topsep=0.5\topsep,itemsep=0ex,parsep=0ex,labelwidth=0.5em,leftmargin=\labelwidth+\labelsep-\itemindent]
\item $\left\langle X_1X_2\cdots X_k\right\rangle_\mathrm{c}$ and $\left\langle Y_1Y_2\cdots Y_l\right\rangle_\mathrm{c}$ contribute each stochastic variable only once to each product,\\
$\circ$ the sum of cumulant orders in each product is $k+l$
\item $\left\langle X_1X_2\cdots X_k\right\rangle_\mathrm{c}$ and $\left\langle Y_1Y_2\cdots Y_l\right\rangle_\mathrm{c}$ contribute one stochastic variable at least to each cumulant in each product,\\
$\circ$ the order of each cumulant is not less than 2,\\
$\circ$ the number of cumulants in each product is not greater than $\min(k,l)$,
\item Each satisfied product appears only once,\\
$\circ$ the numerical coefficient of each product is 1,\\
$\circ$ the number of products with the cumulant orders $i_1,i_2,\cdots,i_r$ is $\left.g_r(k,l;i_1,i_2,\cdots,i_r)\middle/\Delta_{i_1,i_2,\cdots,i_r}\right.$.
\end{itemize}
\end{widetext}

\section{Variances of estimated efficiency-corrected cumulants}

\subsection{Univariate case}

Suppose that there exists a non-negative integer stochastic variable $X$ representing the number of produced objects of interest, such as the multiplicity of particles produced in heavy-ion collisions, which follows the population probability distribution function $P(X)$. The cumulants of $X$ can be generated from
\begin{align}
\left\langle X^k\right\rangle_\mathrm{c}&=\left.\partial_\theta^kG_\mathrm{c}(\theta)\right|_{\theta=0},\\
G_\mathrm{c}(\theta)&=\ln{\sum_{X}{P(X)\mathrm{e}^{\theta X}}}=\ln{\left\langle\mathrm{e}^{\theta X}\right\rangle}.
\end{align}
The $k$th-order cumulant $\left\langle X^k\right\rangle_\mathrm{c}$ can be simply marked by $C_k$.

If the measuring efficiency for each individual object is $\alpha\in[0,1]$, the measured number $x$ follows Eq.~\eqref{eqb1}, and its cumulants can be generated from
\begin{align}
\left\langle x^k\right\rangle_\mathrm{c}&=\left.\partial_\theta^k\tilde{G}_\mathrm{c}(\theta)\right|_{\theta=0},\\
\tilde{G}_\mathrm{c}(\theta)&=\ln{\sum_{x}{\tilde{P}(x)\mathrm{e}^{\theta x}}}=\ln{\left\langle\mathrm{e}^{\theta x}\right\rangle}.
\end{align}

The formulas for efficiency-corrected cumulants up to fourth order derived in Ref.~\cite{Nonaka:2017kko} are presented as
\begin{widetext}
\begin{align}
\langle X\rangle_{\mathrm{c}}&=\frac{1}{\alpha}\langle x\rangle_{\mathrm{c}},\\
\left\langle X^{2}\right\rangle_{\mathrm{c}}&=\frac{1}{\alpha^2}\left\langle x^{2}\right\rangle_{\mathrm{c}}+\left(\frac{1}{\alpha}-\frac{1}{\alpha^2}\right)\langle x\rangle_{\mathrm{c}},\\
\left\langle X^{3}\right\rangle_{\mathrm{c}}&=\frac{1}{\alpha^3}\left\langle x^{3}\right\rangle_{\mathrm{c}}+\left(-\frac{3}{\alpha^3}+\frac{3}{\alpha^2}\right)\left\langle x^{2}\right\rangle_{\mathrm{c}}+\left(\frac{2}{\alpha^3}-\frac{3}{\alpha^2}+\frac{1}{\alpha}\right)\langle x\rangle_{\mathrm{c}},\\
\left\langle X^{4}\right\rangle_{\mathrm{c}}&=\frac{1}{\alpha^4}\left\langle x^{4}\right\rangle_{\mathrm{c}}+\left(-\frac{6}{\alpha^4}+\frac{6}{\alpha^3}\right)\left\langle x^{3}\right\rangle_{\mathrm{c}}+\left(\frac{11}{\alpha^4}-\frac{18}{\alpha^3}+\frac{7}{\alpha^2}\right)\left\langle x^{2}\right\rangle_{\mathrm{c}}+\left(-\frac{6}{\alpha^4}+\frac{12}{\alpha^3}-\frac{7}{\alpha^2}+\frac{1}{\alpha}\right)\langle x\rangle_{\mathrm{c}}.
\end{align}
The formula of the efficiency correction on the right-hand side for the $k$th-order cumulant is represented by $C_k^{\mathrm{corr}}$.

With the covariances of cumulants obtained in the previous section, the variances, as squares of statistical uncertainties, of the estimated efficiency-corrected cumulants are derived accordingly as
\begin{align}
n\cdot\mathrm{Var}\left(\widehat{C}_1^{\mathrm{corr}}\right)&=\frac{1}{\alpha^2}\left\langle x^{2}\right\rangle_{\mathrm{c}},\\
n\cdot\mathrm{Var}\left(\widehat{C}_2^{\mathrm{corr}}\right)&=\frac{1}{\alpha^4}\left(\left\langle x^{4}\right\rangle_{\mathrm{c}}+2\left\langle x^{2}\right\rangle_{\mathrm{c}}^2\right)+\frac{2}{\alpha^2}\left(\frac{1}{\alpha}-\frac{1}{\alpha^2}\right)\left\langle x^{3}\right\rangle_{\mathrm{c}}+\left(\frac{1}{\alpha}-\frac{1}{\alpha^2}\right)^2\left\langle x^{2}\right\rangle_{\mathrm{c}},\\
\begin{split}
n\cdot\mathrm{Var}\left(\widehat{C}_3^{\mathrm{corr}}\right)&=\frac{1}{\alpha^6}\left(\left\langle x^{6}\right\rangle_{\mathrm{c}}+9\left\langle x^{4}\right\rangle_{\mathrm{c}}\left\langle x^{2}\right\rangle_{\mathrm{c}}+9\left\langle x^{3}\right\rangle_{\mathrm{c}}^2+6\left\langle x^{2}\right\rangle_{\mathrm{c}}^3\right)+\frac{2}{\alpha^3}\left(-\frac{3}{\alpha^3}+\frac{3}{\alpha^2}\right)\left(\left\langle x^{5}\right\rangle_{\mathrm{c}}+6\left\langle x^{3}\right\rangle_{\mathrm{c}}\left\langle x^{2}\right\rangle_{\mathrm{c}}\right)\\
&\hphantom{\quad\ }+\left(-\frac{3}{\alpha^3}+\frac{3}{\alpha^2}\right)^2\left(\left\langle x^{4}\right\rangle_{\mathrm{c}}+2\left\langle x^{2}\right\rangle_{\mathrm{c}}^2\right)+\frac{2}{\alpha^3}\left(\frac{2}{\alpha^3}-\frac{3}{\alpha^2}+\frac{1}{\alpha}\right)\left\langle x^{4}\right\rangle_{\mathrm{c}}\\
&\hphantom{\quad\ }+2\left(-\frac{3}{\alpha^3}+\frac{3}{\alpha^2}\right)\left(\frac{2}{\alpha^3}-\frac{3}{\alpha^2}+\frac{1}{\alpha}\right)\left\langle x^{3}\right\rangle_{\mathrm{c}}+\left(\frac{2}{\alpha^3}-\frac{3}{\alpha^2}+\frac{1}{\alpha}\right)^2\left\langle x^{2}\right\rangle_{\mathrm{c}},
\end{split}\\
\begin{split}
n\cdot\mathrm{Var}\left(\widehat{C}_4^{\mathrm{corr}}\right)&=\frac{1}{\alpha^8}\left(\left\langle x^8\right\rangle_\mathrm{c}+16\left\langle x^6\right\rangle_\mathrm{c}\left\langle x^2\right\rangle_\mathrm{c}+48\left\langle x^5\right\rangle_\mathrm{c}\left\langle x^3\right\rangle_\mathrm{c}+34\left\langle x^4\right\rangle_\mathrm{c}^2+72\left\langle x^4\right\rangle_\mathrm{c}\left\langle x^2\right\rangle_\mathrm{c}^2+144\left\langle x^3\right\rangle_\mathrm{c}^2\left\langle x^2\right\rangle_\mathrm{c}+24\left\langle x^2\right\rangle_\mathrm{c}^4\right)\\
&\hphantom{\quad\ }+\frac{2}{\alpha^4}\left(-\frac{6}{\alpha^4}+\frac{6}{\alpha^3}\right)\left(\left\langle x^7\right\rangle_\mathrm{c}+12\left\langle x^5\right\rangle_\mathrm{c}\left\langle x^2\right\rangle_\mathrm{c}+30\left\langle x^4\right\rangle_\mathrm{c}\left\langle x^3\right\rangle_\mathrm{c}+36\left\langle x^3\right\rangle_\mathrm{c}\left\langle x^2\right\rangle_\mathrm{c}^2\right)\\
&\hphantom{\quad\ }+\left(-\frac{6}{\alpha^4}+\frac{6}{\alpha^3}\right)^2\left(\left\langle x^6\right\rangle_\mathrm{c}+9\left\langle x^4\right\rangle_\mathrm{c}\left\langle x^2\right\rangle_\mathrm{c}+9\left\langle x^3\right\rangle_\mathrm{c}^2+6\left\langle x^2\right\rangle_\mathrm{c}^3\right)\\
&\hphantom{\quad\ }+\frac{2}{\alpha^4}\left(\frac{11}{\alpha^4}-\frac{18}{\alpha^3}+\frac{7}{\alpha^2}\right)\left(\left\langle x^6\right\rangle_\mathrm{c}+8\left\langle x^4\right\rangle_\mathrm{c}\left\langle x^2\right\rangle_\mathrm{c}+6\left\langle x^3\right\rangle_\mathrm{c}^2\right)\\
&\hphantom{\quad\ }+2\left(-\frac{6}{\alpha^4}+\frac{6}{\alpha^3}\right)\left(\frac{11}{\alpha^4}-\frac{18}{\alpha^3}+\frac{7}{\alpha^2}\right)\left(\left\langle x^{5}\right\rangle_{\mathrm{c}}+6\left\langle x^{3}\right\rangle_{\mathrm{c}}\left\langle x^{2}\right\rangle_{\mathrm{c}}\right)+\frac{2}{\alpha^4}\left(-\frac{6}{\alpha^4}+\frac{12}{\alpha^3}-\frac{7}{\alpha^2}+\frac{1}{\alpha}\right)\left\langle x^5\right\rangle_\mathrm{c}\\
&\hphantom{\quad\ }+\left(\frac{11}{\alpha^4}-\frac{18}{\alpha^3}+\frac{7}{\alpha^2}\right)^2\left(\left\langle x^{4}\right\rangle_{\mathrm{c}}+2\left\langle x^{2}\right\rangle_{\mathrm{c}}^2\right)+2\left(-\frac{6}{\alpha^4}+\frac{6}{\alpha^3}\right)\left(-\frac{6}{\alpha^4}+\frac{12}{\alpha^3}-\frac{7}{\alpha^2}+\frac{1}{\alpha}\right)\left\langle x^{4}\right\rangle_{\mathrm{c}}
\end{split}\nonumber\\
&\hphantom{\quad\ }+2\left(\frac{11}{\alpha^4}-\frac{18}{\alpha^3}+\frac{7}{\alpha^2}\right)\left(-\frac{6}{\alpha^4}+\frac{12}{\alpha^3}-\frac{7}{\alpha^2}+\frac{1}{\alpha}\right)\left\langle x^{3}\right\rangle_{\mathrm{c}}+\left(-\frac{6}{\alpha^4}+\frac{12}{\alpha^3}-\frac{7}{\alpha^2}+\frac{1}{\alpha}\right)^2\left\langle x^{2}\right\rangle_{\mathrm{c}}.
\end{align}
\end{widetext}

\subsection{Multivariate case}

Suppose that $M$ series of objects of interest are produced with the numbers $\bm{X}=\left(X_1,X_2,\cdots,X_M\right)$ as non-negative integer stochastic variables following the population probability distribution function $P(\bm{X})$. The linear combination of the produced numbers with the numerical coefficients $\bm{a}=\left(a_1,a_2,\cdots,a_M\right)$ is represented by
\begin{equation}
Q_{(\bm{a})}=\bm{a}\cdot\bm{X}=\sum_{i=1}^M{a_iX_i},
\end{equation}
which gives the conserved charge in heavy-ion collisions if $\bm{X}$ and $\bm{a}$ denote the multiplicities and charges of the produced particles of various series, and its cumulants are defined as
\begin{align}
\left\langle Q_{(\bm{a})}^kQ_{(\bm{b})}^l\right\rangle_\mathrm{c}&=\left.\partial_{(\bm{a})}^k\partial_{(\bm{b})}^lG_\mathrm{c}(\bm{\theta})\right|_{\bm{\theta}=0},\\
G_\mathrm{c}(\bm{\theta})&=\ln\sum_{\bm{X}}P(\bm{X})\mathrm{e}^{\bm{\theta}\cdot\bm{X}}=\ln{\left\langle\mathrm{e}^{\bm{\theta}\cdot\bm{X}}\right\rangle},
\end{align}
with
\begin{equation}
\partial_{(\bm{a})}=\sum_{i=1}^M{a_i\partial_{\theta_i}}.
\end{equation}
The $k$th-order diagonal cumulant $\left\langle Q_{(\bm{a})}^k\right\rangle_\mathrm{c}$ can be simply marked by $C_k$.

Consider the measuring efficiency vector $\bm{\alpha}=\left(\alpha_1,\alpha_2,\cdots,\alpha_M\right)\in[0,1]^M$ for these series of objects, and the probability distribution function of the measured numbers $\bm{x}=\left(x_1,x_2,\cdots,x_M\right)$ can be expressed by
\begin{equation}
\tilde{P}(\bm{x})=\sum_{\bm{X}}P(\bm{X})\prod_{i=1}^M\mathcal{B}_{X_i,\alpha_i}\left(x_i\right)
\end{equation}
with $\displaystyle{\alpha_i=\frac{\left\langle x_i\right\rangle}{\langle X_i\rangle}}$.
The linear combination of the measured numbers is represented by
\begin{equation}
q_{(\bm{a})}=\bm{a}\cdot\bm{x},
\end{equation}
and its cumulants are given by
\begin{align}
\left\langle q_{(\bm{a})}^kq_{(\bm{b})}^l\right\rangle_\mathrm{c}&=\left.\partial_{(\bm{a})}^k\partial_{(\bm{b})}^l\tilde{G}_{\mathrm{c}}(\bm{\theta})\right|_{\bm{\theta}=0},\\
\tilde{G}_{\mathrm{c}}(\bm{\theta})&=\ln\sum_{\bm{x}}\tilde{P}(\bm{x})\mathrm{e}^{\bm{\theta}\cdot\bm{x}}=\ln{\left\langle\mathrm{e}^{\bm{\theta}\cdot\bm{x}}\right\rangle}.
\end{align}

The formulas for efficiency-corrected diagonal cumulants up to fourth order are presented as~\cite{Nonaka:2017kko,Kitazawa:2017ljq,Luo:2018ofd,Si:2021mdj}
\begin{widetext}
\begin{align}
\label{eqc1}
\left\langle Q_{(1,0)}\right\rangle_{\mathrm{c}}&=\left\langle q_{(1,1)}\right\rangle_{\mathrm{c}},\\
\label{eqc2}
\left\langle Q_{(1,0)}^{2}\right\rangle_{\mathrm{c}}&=\left\langle q_{(1,1)}^{2}\right\rangle_{\mathrm{c}}+\left\langle q_{(2,1)}\right\rangle_{\mathrm{c}}-\left\langle q_{(2,2)}\right\rangle_{\mathrm{c}},\\
\label{eqc3}
\left\langle Q_{(1,0)}^{3}\right\rangle_{\mathrm{c}}&=\left\langle q_{(1,1)}^{3}\right\rangle_{\mathrm{c}}+3\left\langle q_{(1,1)} q_{(2,1)}\right\rangle_{\mathrm{c}}-3\left\langle q_{(1,1)} q_{(2,2)}\right\rangle_{\mathrm{c}}+\left\langle q_{(3,1)}\right\rangle_{\mathrm{c}}-3\left\langle q_{(3,2)}\right\rangle_{\mathrm{c}}+2\left\langle q_{(3,3)}\right\rangle_{\mathrm{c}},\\
\label{eqc4}
\begin{split}
\left\langle Q_{(1,0)}^{4}\right\rangle_{\mathrm{c}}&=\left\langle q_{(1,1)}^{4}\right\rangle_{\mathrm{c}}+6\left\langle q_{(1,1)}^{2} q_{(2,1)}\right\rangle_{\mathrm{c}}-6\left\langle q_{(1,1)}^{2} q_{(2,2)}\right\rangle_{\mathrm{c}}+4\left\langle q_{(1,1)} q_{(3,1)}\right\rangle_{\mathrm{c}}-12\left\langle q_{(1,1)} q_{(3,2)}\right\rangle_{\mathrm{c}}+8\left\langle q_{(1,1)} q_{(3,3)}\right\rangle_{\mathrm{c}}\\
&\hphantom{\quad\ }+3\left\langle q_{(2,1)}^{2}\right\rangle_{\mathrm{c}}-6\left\langle q_{(2,1)} q_{(2,2)}\right\rangle_{\mathrm{c}}+3\left\langle q_{(2,2)}^{2}\right\rangle_{\mathrm{c}}+\left\langle q_{(4,1)}\right\rangle_{\mathrm{c}}-7\left\langle q_{(4,2)}\right\rangle_{\mathrm{c}}+12\left\langle q_{(4,3)}\right\rangle_{\mathrm{c}}-6\left\langle q_{(4,4)}\right\rangle_{\mathrm{c}},
\end{split}
\end{align}
where
\begin{align}
Q_{(u,v)}&=Q_{\left(\bm{a}^u/{\bm{\alpha}'}^v\right)}=\sum_{i=1}^M{\frac{a_i^u}{{\alpha'}_i^v}X_i},\\
q_{(u,v)}&=q_{\left(\bm{a}^u/{\bm{\alpha}'}^v\right)}=\sum_{i=1}^M{\frac{a_i^u}{{\alpha'}_i^v}x_i}.
\end{align}
Here $\bm{\alpha}'=\left(\alpha'_1,\alpha'_2,\cdots,\alpha'_M\right)\in[0,1]^M$ denotes the correcting efficiency and is not necessarily the same as the realistic measuring efficiency $\bm{\alpha}$. In heavy-ion collisions, we have shown elsewhere that the efficiency correction is valid if $\bm{\alpha}'$ is an average of $\bm{\alpha}$ for the particles with the same charges within each single phase space~\cite{Si:2021mdj}. Similarly in the general case, the correcting efficiency is required to be an average of the measuring efficiency taken respectively in one or several average ranges meeting two requirements:
\begin{itemize}[topsep=0.5\topsep,itemsep=0ex,parsep=0ex,labelwidth=0.5em,leftmargin=\labelwidth+\labelsep-\itemindent]
\item Numerical coefficients of all series of objects in each average range are the same,
\item Object numbers of all series in each average range follow Eq.~\eqref{eqb2}.
\end{itemize}
For each equation in Eqs.~\eqref{eqc1}-\eqref{eqc4}, the left-hand side, which denotes the true value of the produced $k$th-order diagonal cumulant, is represented by $C_k^{\mathrm{true}}$, and the right-hand side as the formula of the efficiency correction with $\bm{\alpha}'$ from the measurement with $\bm{\alpha}$ is marked by $C_k^{\mathrm{corr}}(\bm{\alpha};\bm{\alpha}')$.

With the same technique as the univariate case, the variances of estimated efficiency-corrected cumulants up to second order can be obtained accordingly as
\begin{align}
\label{eqe1}
n\cdot\mathrm{Var}\left(\widehat{C}_1^{\mathrm{corr}}\right)&=\left\langle q_{(1,1)}^2\right\rangle_{\mathrm{c}},\\
\label{eqe2}
\begin{split}
n\cdot\mathrm{Var}\left(\widehat{C}_2^{\mathrm{corr}}\right)&=\left\langle q_{(1,1)}^{4}\right\rangle_{\mathrm{c}}+2\left\langle q_{(1,1)}^{2}\right\rangle_{\mathrm{c}}^{2}+2\left\langle q_{(1,1)}^{2} q_{(2,1)}\right\rangle_{\mathrm{c}}-2\left\langle q_{(1,1)}^{2} q_{(2,2)}\right\rangle_{\mathrm{c}}\\
&\hphantom{\quad\ }+\left\langle q_{(2,1)}^{2}\right\rangle_{\mathrm{c}}-2\left\langle q_{(2,1)} q_{(2,2)}\right\rangle_{\mathrm{c}}+\left\langle q_{(2,2)}^{2}\right\rangle_{\mathrm{c}},
\end{split}
\end{align}
and the results for third- and fourth-order cumulants can be found in Appendix~\ref{appendix}. In addition, the covariance between the first two order ones is derived as
\begin{equation}
\label{eqer1}
n\cdot\mathrm{Cov}\left(\widehat{C}_2^{\mathrm{corr}},\widehat{C}_1^{\mathrm{corr}}\right)=\left\langle q_{(1,1)}^3\right\rangle_{\mathrm{c}}+\left\langle q_{(1,1)}q_{(2,1)}\right\rangle_{\mathrm{c}}-\left\langle q_{(1,1)}q_{(2,2)}\right\rangle_{\mathrm{c}},
\end{equation}
and some other covariances are also presented in Appendix~\ref{appendix}. In fluctuation analysis in heavy-ion collision experiments, cumulant ratios, such as $\left.C_2\middle/C_1\right.$, $\left.C_3\middle/C_2\right.$ and $\left.C_4\middle/C_2\right.$, are commonly studied, since the fluctuations induced by the system volume can be canceled~\cite{Luo:2017faz}, and their statistical uncertainties can be estimated from the variances and covariances in the above derivations by
\begin{equation}
\label{eqer}
\mathrm{Var}\left(\frac{\widehat{C}_k^{\mathrm{corr}}}{\widehat{C}_l^{\mathrm{corr}}}\right)=\left(\frac{C_k^{\mathrm{corr}}}{C_l^{\mathrm{corr}}}\right)^2\left(\frac{\mathrm{Var}\left(\widehat{C}_k^{\mathrm{corr}}\right)}{\left(C_k^{\mathrm{corr}}\right)^2}+\frac{\mathrm{Var}\left(\widehat{C}_l^{\mathrm{corr}}\right)}{\left(C_l^{\mathrm{corr}}\right)^2}-\frac{2\mathrm{Cov}\left(\widehat{C}_k^{\mathrm{corr}},\widehat{C}_l^{\mathrm{corr}}\right)}{C_k^{\mathrm{corr}}C_l^{\mathrm{corr}}}\right).
\end{equation}
\end{widetext}

\subsection{Components of variances in multivariate case}

In Ref.~\cite{Si:2021mdj}, statistical fluctuations of efficiency-corrected higher-order cumulants are observed to depend on the non-uniformity of the valid correcting efficiency in a Monte Carlo simulation. The components of variances are fully decomposed and discussed via an analytical derivation in the following to understand how the statistical uncertainties are affected.

Suppose that all series of the produced objects in the previous section are in a single phase space, which requires that the object numbers follow Eq.~\eqref{eqb2} with the total number $X$ and the multinomial probability vector $\bm{p}$~\cite{Si:2021mdj}. The cumulants and factorial cumulants of
\begin{equation}
Q_{(a)}=aX
\end{equation}
are generated by
\begin{align}
\left\langle Q_{(a)}^kQ_{(b)}^l\right\rangle_\mathrm{c}&=\left.\partial_{(a)}^k\partial_{(b)}^lG'_\mathrm{c}(\theta)\right|_{\theta=0},\\
G'_\mathrm{c}(\theta)&=\ln\sum_{X}P'(X)\mathrm{e}^{\theta X}=\ln\left\langle\mathrm{e}^{\theta X}\right\rangle,\\
\left\langle Q_{(a)}^kQ_{(b)}^l\right\rangle_\mathrm{fc}&=\left.\bar\partial_{(a)}^k\bar\partial_{(b)}^lG'_\mathrm{fc}(s)\right|_{s=1},\\
G'_\mathrm{fc}(s)&=\ln\sum_{X}P'(X)s^X=\ln\left\langle s^X\right\rangle,
\end{align}
respectively, with
\begin{align}
\partial_{(a)}&=a\partial_\theta,\\
\bar\partial_{(a)}&=a\partial_s.
\end{align}

In addition, the factorial cumulants of $\bm{X}$ and $\bm{x}$ are defined by
\begin{align}
\left\langle Q_{(\bm{a})}^kQ_{(\bm{b})}^l\right\rangle_\mathrm{fc}&=\left.\bar{\partial}_{(\bm{a})}^k\bar{\partial}_{(\bm{b})}^lG_\mathrm{fc}(\bm{s})\right|_{\bm{s}=1},\\
G_\mathrm{fc}(\bm{s})&=\ln\sum_{\bm{X}}P(\bm{X})\prod_{i=1}^Ms_i^{X_i}=\ln{\left\langle\prod_{i=1}^M{s_i^{X_i}}\right\rangle},\\
\left\langle q_{(\bm{a})}^kq_{(\bm{b})}^l\right\rangle_\mathrm{fc}&=\left.\bar\partial_{(\bm{a})}^k\bar\partial_{(\bm{b})}^l\tilde{G}_{\mathrm{fc}}(\bm{s})\right|_{\bm{s}=1}\\
\tilde{G}_{\mathrm{fc}}(\bm{s})&=\ln\sum_{\bm{x}}\tilde{P}(\bm{x})\prod_{i=1}^Ms_i^{x_i}=\ln{\left\langle\prod_{i=1}^M{s_i^{x_i}}\right\rangle},
\end{align}
respectively, with
\begin{equation}
\bar{\partial}_{(\bm{a})}=\sum_{i=1}^M{a_i\partial_{s_i}}.
\end{equation}
Relations between cumulants and factorial cumulants have been derived in Ref.~\cite{Nonaka:2017kko}.

$G_\mathrm{fc}(\bm{s})$ can be converted into $G'_\mathrm{fc}(s)$ with~\cite{Si:2021mdj}
\begin{equation}
\begin{split}
G_\mathrm{fc}(\bm{s})&=\ln\sum_{X}P'(X)\sum_{\bm{X}}\mathcal{M}_{X,\bm{p}}(\bm{X})\prod_{i=1}^Ms_i^{X_i}\\
&=\ln\sum_{X}P'(X)\left(\sum_{i=1}^Mp_is_i\right)^X\\
&=G'_\mathrm{fc}\left(\sum_{i=1}^Mp_is_i\right),
\end{split}
\end{equation}
where the second line is obtained by the multinomial expansion, so
\begin{equation}
\bar\partial_{(\bm{a})}^k\bar\partial_{(\bm{b})}^lG_\mathrm{fc}(\bm{s})=\bar\partial_{(\bm{a}\cdot\bm{p})}^k\bar\partial_{(\bm{b}\cdot\bm{p})}^lG'_\mathrm{fc}(s).
\end{equation}

The relation between $\tilde{G}_\mathrm{fc}(\bm{s})$ and $G_\mathrm{fc}(\bm{s})$ can be obtained by~\cite{Nonaka:2017kko,Kitazawa:2017ljq}
\begin{equation}
\begin{split}
\tilde{G}_\mathrm{fc}(\bm{s})&=\ln\sum_{\bm{X}}P(\bm{X})\sum_{\bm{x}}\prod_{i=1}^M\mathcal{B}_{X_i,\alpha_i}\left(x_i\right)s_i^{x_i}\\
&=\ln\sum_{\bm{X}}P(\bm{X})\prod_{i=1}^M\left(\alpha_is_i+\left(1-\alpha_i\right)\right)^{X_i}\\
&=G_\mathrm{fc}(\bm{s}'),
\end{split}
\end{equation}
where $s'_i=\alpha_is_i+\left(1-\alpha_i\right)$ ($i$ = 1, 2, $\cdots$, $M$), so
\begin{equation}
\bar\partial_{(\bm{a})}^k\bar\partial_{(\bm{b})}^l\tilde{G}_\mathrm{fc}(\bm{s})=\bar\partial_{(\bm{a}\bm{\alpha})}^k\bar\partial_{(\bm{b}\bm{\alpha})}^lG_\mathrm{fc}(\bm{s}),
\end{equation}
which connects the produced and measured factorial cumulants. Here we define the expression
\begin{equation}
\bar\partial_{(\bm{a}\bm{\alpha}/\bm{\alpha}')}=\sum_{i=1}^M\frac{a_i\alpha_i}{\alpha'_i}\partial_{s_i}.
\end{equation}

In a valid efficiency correction employing
\begin{equation}
\bm{\alpha}'=\left(\bar{\alpha},\cdots,\bar{\alpha},\alpha_{m+1},\cdots,\alpha_M\right),
\end{equation}
with
\begin{equation}
\bar{\alpha}=\frac{\displaystyle{\sum_{i=1}^{m}\left\langle x_i\right\rangle}}{\displaystyle{\sum_{i=1}^{m}\left\langle X_i\right\rangle}}=\frac{\displaystyle{\sum_{i=1}^{m}\alpha_i\left\langle X_i\right\rangle}}{\displaystyle{\sum_{i=1}^{m}\left\langle X_i\right\rangle}}=\frac{\displaystyle{\sum_{i=1}^{m}\alpha_ip_i}}{\displaystyle{\sum_{i=1}^{m}p_i}}
\end{equation}
as the averaged efficiency with an average range covering the first $m\in[1,M]$ series of objects, it is required that numerical coefficients of objects in the average range should be the same; that is, $a_1=a_2=\cdots=a_m$~\cite{Si:2021mdj}. The relations between generating functions of factorial cumulants can be derived as
\begin{equation}
\begin{split}
&\hphantom{\quad\ }\bar\partial_{(\bm{a}^u/\bm{\alpha}'^v)}\tilde{G}_\mathrm{fc}(\bm{s})\\
&=\bar\partial_{(\bm{a}^u\bm{\alpha}/\bm{\alpha}'^v)}G_\mathrm{fc}(\bm{s})\\
&=\left(\frac{a_1^u}{\bar{\alpha}^v}\sum_{i=1}^m{\alpha_ip_i}+\sum_{i=m+1}^M\frac{a_i^u\alpha_i}{\alpha_i^v}p_i\right)\partial_sG'_\mathrm{fc}(s)\\
&=\left(\frac{a_1^u}{\bar{\alpha}^{v-1}}\sum_{i=1}^m{p_i}+\sum_{i=m+1}^M\frac{a_i^u}{\alpha_i^{v-1}}p_i\right)\partial_sG'_\mathrm{fc}(s)\\
&=\bar\partial_{(\bm{a}^u/\bm{\alpha}'^{v-1})}G_\mathrm{fc}(\bm{s}),
\end{split}
\end{equation}
and so forth,
\begin{equation}
\left\langle q_{(u,v)}^kq_{(u',v')}^l\right\rangle_\mathrm{fc}=\left\langle Q_{(u,v-1)}^kQ_{(u',v'-1)}^l\right\rangle_\mathrm{fc}.
\end{equation}

Hence, the variance of the estimated efficiency-corrected first-order cumulant can be decomposed as
\begin{align}
\begin{split}
n\cdot\mathrm{Var}\left(\widehat{C}_1^{\mathrm{corr}}\right)&=\left\langle q_{(1,1)}^2\right\rangle_{\mathrm{c}}\\
&=\left\langle q_{(1,1)}^2\right\rangle_{\mathrm{fc}}+\left\langle q_{(2,2)}\right\rangle_{\mathrm{fc}}\\
&=\left\langle Q_{(1,0)}^2\right\rangle_{\mathrm{fc}}+\left\langle Q_{(2,1)}\right\rangle_{\mathrm{fc}}\\
&=\left\langle Q_{(1,0)}^2\right\rangle_{\mathrm{c}}-\left\langle Q_{(2,0)}\right\rangle_{\mathrm{fc}}+\left\langle Q_{(2,1)}\right\rangle_{\mathrm{fc}}\\
&=n\cdot\mathrm{Var}\left(\widehat{C}_1^{\mathrm{true}}\right)+\left(\left\langle Q_{(2,1)}\right\rangle_{\mathrm{fc}}-\left\langle Q_{(2,0)}\right\rangle_{\mathrm{fc}}\right),
\end{split}\nonumber
\end{align}
with two additive parts: one is an intrinsic part only dependent on the initial population fluctuations, and the other also depends on the correcting efficiency. One can find that
\begin{equation}
n\cdot\mathrm{Var}\left(\widehat{C}_1^{\mathrm{corr}}\right)\geq n\cdot\mathrm{Var}\left(\widehat{C}_1^{\mathrm{true}}\right),
\end{equation}
from the expression of the second part
\begin{equation}
\label{eqfc1}
\begin{split}
&\hphantom{\quad\ }\left\langle Q_{(2,1)}\right\rangle_{\mathrm{fc}}-\left\langle Q_{(2,0)}\right\rangle_{\mathrm{fc}}\\
&=\left\langle Q_{(\bm{a}^2/\bm{\alpha}')}\right\rangle_{\mathrm{fc}}-\left\langle Q_{(\bm{a}^2)}\right\rangle_{\mathrm{fc}}\\
&=\left.\bar\partial_{(\bm{a}^2/\bm{\alpha}')}G_\mathrm{fc}(\bm{s})\right|_{\bm{s}=1}-\left.\bar\partial_{(\bm{a}^2)}G_\mathrm{fc}(\bm{s})\right|_{\bm{s}=1}\\
&=\left.\bar\partial_{(\bm{a}^2(1/\bm{\alpha}'-1))}G_\mathrm{fc}(\bm{s})\right|_{\bm{s}=1}\\
&\geq0,
\end{split}
\end{equation}
since the numerical coefficients $\bm{a}^2(1/\bm{\alpha}'-1)$ is always non-negative for each series of objects. The second part is not explicitly related to the realistic measuring efficiency $\bm{\alpha}$ but its averaged value, since $\bm{\alpha}'$ keeps the same averaged value as $\bm{\alpha}$. It is obvious that the second part of the variance decreases to zero as the efficiency increases to unity. To study the contribution of non-uniformity of the correcting efficiency, the efficiency-dependent part can be further decomposed with its generating function as
\begin{equation}
\left\langle Q_{(\bm{a}^2/\bm{\alpha}')}\right\rangle_{\mathrm{fc}}=\left.\left(\frac{a_1^2}{\bar{\alpha}}\sum_{i=1}^m{p_i}+\sum_{i=m+1}^M{\frac{a_i^2}{\alpha_i}p_i}\right)\partial_sG'_{\mathrm{fc}}(s)\right|_{s=1}.
\end{equation}
With Cauchy's inequality, it can be derived that
\begin{equation}
\frac{1}{\bar{\alpha}}\sum_{i=1}^m{p_i}=\left.\left(\sum_{i=1}^m{p_i}\right)^2\middle/\sum_{i=1}^m{\alpha_ip_i}\right.\leq\sum_{i=1}^m{\frac{1}{\alpha_i}p_i},
\end{equation}
in other words,
\begin{equation}
\left\langle Q_{(\bm{a}^2/\bm{\alpha}')}\right\rangle_{\mathrm{fc}}\leq\left\langle Q_{(\bm{a}^2/\bm{\alpha})}\right\rangle_{\mathrm{fc}},
\end{equation}
and the relationship between the variances of estimated true or efficiency-corrected cumulants can be obtained as
\begin{widetext}
\begin{equation}
\mathrm{Var}\left(\widehat{C}_1^{\mathrm{true}}\right)\leq\mathrm{Var}\left(\widehat{C}_1^{\mathrm{corr}}(\bm{\alpha};\bar{\bm{\alpha}})\right)\leq\mathrm{Var}\left(\widehat{C}_1^{\mathrm{corr}}(\bm{\alpha};\bm{\alpha}')\right)\leq\mathrm{Var}\left(\widehat{C}_1^{\mathrm{corr}}(\bm{\alpha};\bm{\alpha})\right),
\end{equation}
where $\bar{\bm{\alpha}}$ denotes the $\bm{\alpha}'$ with $m=M$ as the completely averaged efficiency without any non-uniformity in the single phase space. The averaged efficiency correction can suppress the statistical uncertainties of higher-order cumulants compared with use of a realistic efficiency correction, and the variance decreases as the average range of efficiency expands, since the non-uniformity of efficiency does not contribute to the fluctuations.

Similarly, the variance of the estimated efficiency-corrected second-order cumulant can be decomposed as
\begin{equation}
\begin{split}
n\cdot\mathrm{Var}\left(\widehat{C}_2^{\mathrm{corr}}\right)&=n\cdot\mathrm{Var}\left(\widehat{C}_2^{\mathrm{true}}\right)+4\left\langle Q_{(1,0)}^{2}Q_{(2,1)}\right\rangle_{\mathrm{fc}}-4\left\langle Q_{(1,0)}^{2} Q_{(2,0)}\right\rangle_{\mathrm{fc}}\\
&\hphantom{\quad\ }+4\left\langle Q_{(1,0)}^{2}\right\rangle_{\mathrm{fc}}\left\langle Q_{(2,1)}\right\rangle_{\mathrm{fc}}-4\left\langle Q_{(1,0)}^{2}\right\rangle_{\mathrm{fc}}\left\langle Q_{(2,0)}\right\rangle_{\mathrm{fc}}+4\left\langle Q_{(1,0)} Q_{(3,1)}\right\rangle_{\mathrm{fc}}-4\left\langle Q_{(1,0)} Q_{(3,0)}\right\rangle_{\mathrm{fc}}\\
&\hphantom{\quad\ }+2\left\langle Q_{(2,1)}^{2}\right\rangle_{\mathrm{fc}}-2\left\langle Q_{(2,0)}^{2}\right\rangle_{\mathrm{fc}}+2\left\langle Q_{(2,1)}\right\rangle_{\mathrm{fc}}^{2}-2\left\langle Q_{(2,0)}\right\rangle_{\mathrm{fc}}^{2}+\left\langle Q_{(4,1)}\right\rangle_{\mathrm{fc}}-\left\langle Q_{(4,0)}\right\rangle_{\mathrm{fc}},
\end{split}
\end{equation}
with two additive parts following the same principle as the first-order case. One can also find that
\begin{equation}
n\cdot\mathrm{Var}\left(\widehat{C}_2^{\mathrm{corr}}\right)\geq n\cdot\mathrm{Var}\left(\widehat{C}_2^{\mathrm{true}}\right),
\end{equation}
since the second part is always non-negative, which can be proved by expressing it similarly with Eq.~\eqref{eqfc1} with non-negative numerical coefficients.
\end{widetext}

\section{Toy model analysis}

In this section, a toy Monte Carlo model is employed with several sets of efficiency with various non-uniformity for particle multiplicity fluctuations in heavy-ion collisions to check the analytical derivation in the previous section.

For the Monte Carlo production of the data sample, $10^7$ events were produced with the multiplicities of positively ($N_\mathrm{pos}$) and negatively ($N_\mathrm{neg}$) charged particles with charges 1 and $-$1 following two independent Poisson distributions with parameters $\lambda_1$ = 12 and $\lambda_2$ = 8, respectively. Thus, the produced net-charge $N_\mathrm{net}=N_\mathrm{pos}-N_\mathrm{neg}$ follows the Skellam distribution with parameters $\left(\lambda_1,\lambda_2\right)$, whose population odd- and even-order cumulants are $\lambda_1-\lambda_2$ and $\lambda_1+\lambda_2$, respectively. Note that these particles can be considered in a single phase space since their multiplicities follow Eq.~\eqref{eqb2}.

Each of the particles in the produced events is independently allocated two parameters $p_\mathrm{T}$ and $\varphi$ sampled from the probability distribution functions
\begin{align}
f\left(p_\mathrm{T}\right)&\sim p_\mathrm{T}\mathrm{exp}\left(-p_\mathrm{T}\middle/t\right),0.4\le p_\mathrm{T}<2,\\
g(\varphi)&\sim\mathrm{Uniform}(0, 2\pi),
\end{align}
where $t$ = 0.26 and 0.22 for positively and negatively charged particles, respectively.

The efficiency is considered as a two-dimensional function of the particle $p_\mathrm{T}$ and $\varphi$, which consists of two independent one-dimensional components defined as
\begin{equation}
u\left(p_\mathrm{T}\right)=p_0\mathrm{exp}\left(-\left(p_1\middle/p_\mathrm{T}\right)^{p_2}\right),
\end{equation}
where $\left(p_0,p_1,p_2\right)$ = (0.7, 0.4, 4.8) for positively and (0.6, 0.4, 4.2) for negatively
charged particles, respectively, and
\begin{equation}
v(\varphi)=\left\{\begin{alignedat}{2}
&1-p_3\left(p_4-1\right),&k\le\left.\varphi\middle/\dfrac{\pi}{6}\right.<k+\dfrac{1}{p_4},\\
&1+p_3,&k+\dfrac{1}{p_4}\le\left.\varphi\middle/\dfrac{\pi}{6}\right.<k+1,
\end{alignedat}\right.(k\in\mathbb{Z})
\end{equation}
where $\left(p_3,p_4\right)$ = (0.4, 3). Their averaged values weighted by the input particle $p_\mathrm{T}$ or $\varphi$ distribution are defined as
\begin{align}
\mu\left[u\left(p_\mathrm{T}\right)\right]&=\dfrac{\displaystyle{\int_{0.4}^2f\left(p_\mathrm{T}\right)u\left(p_\mathrm{T}\right)\mathrm{d}p_\mathrm{T}}}{\displaystyle{\int_{0.4}^2f\left(p_\mathrm{T}\right)\mathrm{d}p_\mathrm{T}}},\\
\mu\left[v(\varphi)\right]&=\dfrac{\displaystyle{\int_0^{2\pi}g(\varphi)v(\varphi)\mathrm{d}\varphi}}{\displaystyle{\int_0^{2\pi}g(\varphi)\mathrm{d}\varphi}}=1,
\end{align}
respectively.

Three sets of efficiency with different non-uniformity are given by
\begin{align}
\label{eqt1}
\varepsilon_0\left(p_\mathrm{T},\varphi\right)&=u\left(p_\mathrm{T}\right)v(\varphi),\\
\label{eqt2}
\varepsilon_1\left(p_\mathrm{T},\varphi\right)&=u\left(p_\mathrm{T}\right),\\
\label{eqt3}
\varepsilon_2\left(p_\mathrm{T},\varphi\right)&=\mu\left[u\left(p_\mathrm{T}\right)\right],
\end{align}
whose averaged values weighted by the input particle distribution are equal to $\mu\left[u\left(p_\mathrm{T}\right)\right]$. The non-uniformity of $\varepsilon_i$ ($i$ = 0, 1, 2) decreases monotonically as $i$ increases, and $\varepsilon_2$ is completely uniform. Note that $\varepsilon_1$ is an average of $\varepsilon_0$, and $\varepsilon_2$ is an average of $\varepsilon_{0,1}$ for particles with the same charges within a single phase-space distribution.

\begin{figure}[htbp]
\centering
\includegraphics[width=0.8\linewidth]{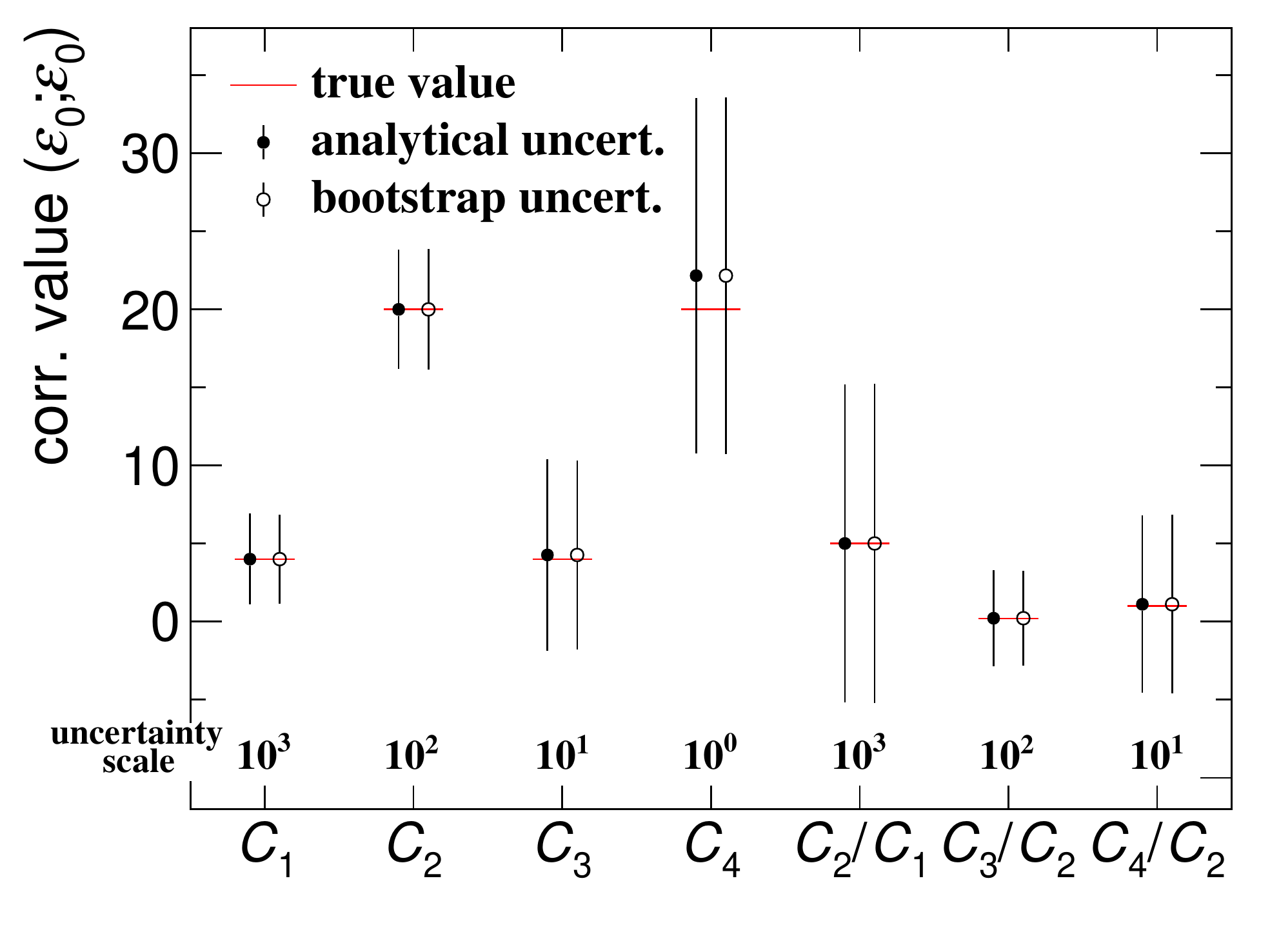}
\caption{(Color online) The $C_k^\mathrm{corr}$ ($k$ = 1, 2, 3, 4) and the ratios $\left.C_2^\mathrm{corr}\middle/C_1^\mathrm{corr}\right.$, $\left.C_3^\mathrm{corr}\middle/C_2^\mathrm{corr}\right.$ and $\left.C_4^\mathrm{corr}\middle/C_2^\mathrm{corr}\right.$ measured and corrected with $\varepsilon_0$ defined in Eq.~\eqref{eqt1}. Uncertainties of solid and open circles are obtained from the analytical and bootstrap methods, respectively. Uncertainty bars are magnified with the scales shown at the bottom to be visible.}
\label{fig1}
\end{figure}

\begin{figure*}[htbp]
\centering
\includegraphics[width=\textwidth]{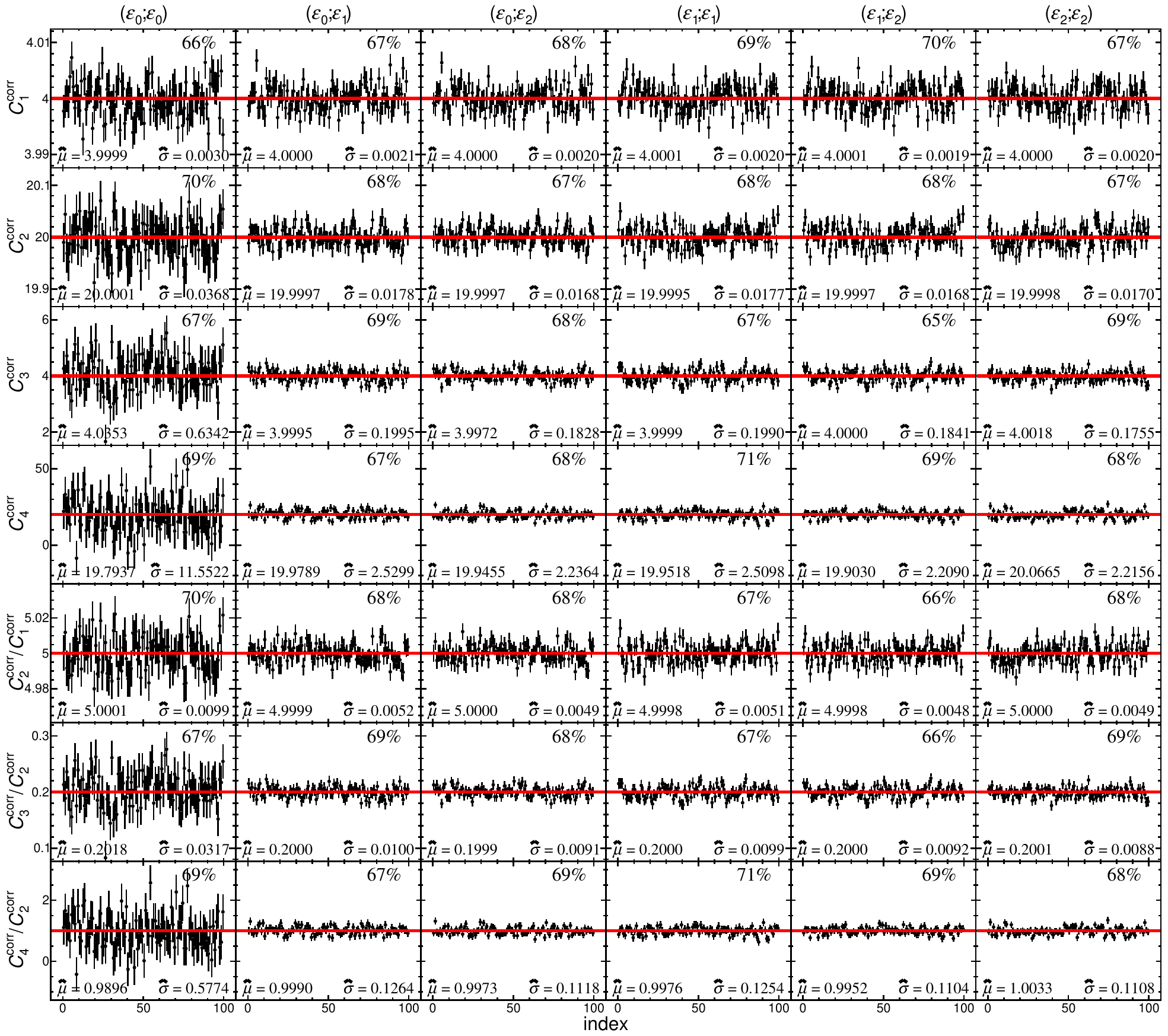}
\caption{(Color online) The $C_k^\mathrm{corr}$ ($k$ = 1, 2, 3, 4) and the ratios $\left.C_2^\mathrm{corr}\middle/C_1^\mathrm{corr}\right.$, $\left.C_3^\mathrm{corr}\middle/C_2^\mathrm{corr}\right.$ and $\left.C_4^\mathrm{corr}\middle/C_2^\mathrm{corr}\right.$ measured with $\varepsilon_i$ and corrected with $\varepsilon_j$ ($i$, $j$ = 0, 1, 2 and $i\leq j$) defined in Eqs.~\eqref{eqt1}-\eqref{eqt3}. Each panel shows 100 points instead of 1000. The red solid lines denote the true values of cumulants or their ratios. The number at the top right of each panel represents the fraction of the uncertainty bars of points touching the red solid lines. The $\hat{\mu}$ and $\hat{\sigma}$ at the bottom of each panel show the mean value and the standard deviation of 1000 points, respectively.}
\label{fig2}
\end{figure*}

For the simulated data sample, each produced particle is sampled with $\varepsilon_i$ ($i$ = 0, 1, 2) defined in Eqs.~\eqref{eqt1}-\eqref{eqt3} as the probability. The so-called track-by-track efficiency correction~\cite{Luo:2018ofd} with Eqs.~\eqref{eqc1}-\eqref{eqc4} employing $\varepsilon_j$ ($j$ = 0, 1, 2 and $j\geq i$) is performed for each of the three measurements to obtain the $N_\mathrm{net}$ $C_k^\mathrm{corr}\left(\varepsilon_i;\varepsilon_j\right)$ ($k$ = 1, 2, 3, 4) and their ratios, all of which can be successfully applied due to the validity of the averaged efficiency correction~\cite{Si:2021mdj}. The statistical uncertainties are analytically estimated from Eqs.~\eqref{eqe1}-\eqref{eqer}~and~\eqref{eqe3}-\eqref{eqer4}. For example, the $C_k^\mathrm{corr}$ ($k$ = 1, 2, 3, 4) and the ratios $\left.C_2^\mathrm{corr}\middle/C_1^\mathrm{corr}\right.$, $\left.C_3^\mathrm{corr}\middle/C_2^\mathrm{corr}\right.$ and $\left.C_4^\mathrm{corr}\middle/C_2^\mathrm{corr}\right.$ measured and corrected with $\varepsilon_0$ are shown in Fig.~\ref{fig1}, whose uncertainty bars are magnified by the factor shown on the bottom scale in order to make them visible. The uncertainties obtained from the bootstrap method~\cite{Luo:2018ofd} with 300 resamplings are also shown for comparison, and little difference between uncertainties from the two different methods can be observed, which supports the validity of the analytical method of statistical uncertainty estimation for efficiency-corrected higher-order cumulants. The bootstrap method is a computationally intensive method whose CPU time cost is proportional to the number of resampling times, number of events and number of tracks of the data in heavy-ion collision experiments. Based on this Mont Carlo sample, the bootstrap method costs about $1.2\times 10^5$ seconds on the CPU ``Intel Xeon E5-2650 v2 @ 2.60 GHz'', however, the analytical method costs only 0.1 second on the same CPU, which improves the computational efficiency more than $10^6$ times.

The above procedures, including the production, measurement and efficiency correction, are repeated 1000 times independently. The cumulants up to fourth order measured with $\varepsilon_i$ and corrected with $\varepsilon_j$ ($i$, $j$ = 0, 1, 2 and $i\leq j$) defined in Eqs.~\eqref{eqt1}-\eqref{eqt3} are shown in Fig.~\ref{fig2} with 100 points instead of 1000 in each panel. The $\hat{\mu}$ and $\hat{\sigma}$ shown at the bottom of each panel represent the mean value and the standard deviation of 1000 points, respectively. The red solid lines denote the true values of cumulants, and the fraction of the uncertainty bars of efficiency-corrected cumulants touching the red solid lines is shown at the top right of each panel. All of the fractions are observed around 68\% which is consistent with the 1-$\sigma$ probability of the Gaussian distribution, which strongly supports the analytical statistical uncertainty estimation for higher-order cumulants while considering efficiency corrections with various non-uniformity. It can also be observed that the statistical fluctuations of efficiency-corrected cumulants quantified by $\hat{\sigma}$ are not obviously affected by the non-uniformity of the realistic measuring efficiency but depend on the non-uniformity of the correcting efficiency when fixing one and tuning another, which supports the uncertainty component study in the previous section. Through tuning the non-uniformity of the correcting efficiency, the fluctuations of $C_1^\mathrm{corr}$, $C_2^\mathrm{corr}$ and $\left.C_2^\mathrm{corr}\middle/C_1^\mathrm{corr}\right.$ are less variable than those of $C_3^\mathrm{corr}$, $C_4^\mathrm{corr}$, $\left.C_3^\mathrm{corr}\middle/C_2^\mathrm{corr}\right.$ and $\left.C_4^\mathrm{corr}\middle/C_2^\mathrm{corr}\right.$, which shows that the statistical uncertainties of higher-order cumulants are more sensitive to the non-uniformity of the correcting efficiency, so the statistical uncertainty suppression of the valid averaged efficiency correction is stronger for higher-order cumulants.

\section{Summary}

In this paper, for the first time, the analytical formulas of the statistical uncertainty estimation of efficiency-corrected higher-order cumulants both in univariate and multivariate cases are derived. The statistical uncertainties are decomposed into essential components, and it is mathematically shown that they correlate to the initial population fluctuations and also depend on the averaged value of the efficiency in the realistic measurement and non-uniformity of the efficiency employed in the valid efficiency correction. A larger averaged value or less non-uniformity of efficiency results in smaller statistical uncertainties for the cumulants. As an application in heavy-ion collisions, a toy Monte Carlo model analysis of higher-order cumulants for particle multiplicity fluctuations with efficiency corrections is found to be consistent with the mathematical results. The statistical uncertainties estimated via the analytical method are identical compared with that from the bootstrap method but with $10^6$ times faster in terms of CPU consumption and are independent on the size of the Monte Carlo sample. It is also observed that the statistical uncertainties of higher-order cumulants are more sensitive to the non-uniformity of the efficiency employed in the efficiency correction and more strongly suppressed by the valid averaged efficiency correction. This analysis can be used to estimate and reduce the statistical uncertainties of efficiency-corrected higher-order cumulants, which is crucial to fluctuation research in many fields besides the search for the QCD critical point in heavy-ion collisions.

\acknowledgements

This work was supported by National Natural Science Foundation of China with Grant No.~11890712, National Key Research and Development Program of China with Grant No.~2018YFE0205200 and 2018YFE0104700, Strategic Priority Research Program of Chinese Academy of Sciences with Grant No.~XDB34000000, Anhui Provincial Natural Science Foundation with Grant No.~1808085J02, and Innovation Fund of Key Laboratory of Quark and Lepton Physics with Grant No.~QLPL2020P01.

\appendix

\begin{widetext}
\section{}
\label{appendix}

The variances of estimated efficiency-corrected third- and fourth-order cumulants in the multivariate case are shown below.
\begin{align}
\label{eqe3}
\begin{split}
n\cdot\mathrm{Var}\left(\widehat{C}_3^{\mathrm{corr}}\right)&=\left\langle q_{(1,1)}^{6}\right\rangle_{\mathrm{c}}+9\left\langle q_{(1,1)}^{4}\right\rangle\left\langle q_{(1,1)}^{2}\right\rangle_{\mathrm{c}}+9\left\langle q_{(1,1)}^{3}\right\rangle_{\mathrm{c}}^{2}+6\left\langle q_{(1,1)}^{2}\right\rangle_{\mathrm{c}}^{3}\\
&\hphantom{\quad\ }+9\left(\left\langle q_{(1,1)}^{2} q_{(2,1)}^{2}\right\rangle_{\mathrm{c}}+\left\langle q_{(1,1)}^{2}\right\rangle_{\mathrm{c}}\left\langle q_{(2,1)}^{2}\right\rangle_{\mathrm{c}}+\left\langle q_{(1,1)} q_{(2,1)}\right\rangle_{\mathrm{c}}^{2}\right)\\
&\hphantom{\quad\ }+9\left(\left\langle q_{(1,1)}^{2} q_{(2,2)}^{2}\right\rangle_{\mathrm{c}}+\left\langle q_{(1,1)}^{2}\right\rangle_{\mathrm{c}}\left\langle q_{(2,2)}^{2}\right\rangle_{\mathrm{c}}+\left\langle q_{(1,1)} q_{(2,2)}\right\rangle_{\mathrm{c}}^{2}\right)\\
&\hphantom{\quad\ }+\left\langle q_{(3,1)}^{2}\right\rangle_{\mathrm{c}}+9\left\langle q_{(3,2)}^{2}\right\rangle_{\mathrm{c}}+4\left\langle q_{(3,3)}^{2}\right\rangle_{\mathrm{c}}\\
&\hphantom{\quad\ }+6\left(\left\langle q_{(1,1)}^{4} q_{(2,1)}\right\rangle_{\mathrm{c}}+3\left\langle q_{(1,1)}^{3}\right\rangle_{\mathrm{c}}\left\langle q_{(1,1)} q_{(2,1)}\right\rangle_{\mathrm{c}}+3\left\langle q_{(1,1)}^{2} q_{(2,1)}\right\rangle_{\mathrm{c}}\left\langle q_{(1,1)}^{2}\right\rangle_{\mathrm{c}}\right)\\
&\hphantom{\quad\ }-6\left(\left\langle q_{(1,1)}^{4} q_{(2,2)}\right\rangle_{\mathrm{c}}+3\left\langle q_{(1,1)}^{3}\right\rangle_{\mathrm{c}}\left\langle q_{(1,1)} q_{(2,2)}\right\rangle_{\mathrm{c}}+3\left\langle q_{(1,1)}^{2} q_{(2,2)}\right\rangle_{\mathrm{c}}\left\langle q_{(1,1)}^{2}\right\rangle_{\mathrm{c}}\right)\\
&\hphantom{\quad\ }+2\left\langle q_{(1,1)}^{3} q_{(3,1)}\right\rangle_{\mathrm{c}}-6\left\langle q_{(1,1)}^{3} q_{(3,2)}\right\rangle_{\mathrm{c}}+4\left\langle q_{(1,1)}^{3} q_{(3,3)}\right\rangle_{\mathrm{c}}\\
&\hphantom{\quad\ }-18\left(\left\langle q_{(1,1)}^{2} q_{(2,1)} q_{(2,2)}\right\rangle_{\mathrm{c}}+\left\langle q_{(1,1)}^{2}\right\rangle_{\mathrm{c}}\left\langle q_{(2,1)} q_{(2,2)}\right\rangle_{\mathrm{c}}+\left\langle q_{(1,1)} q_{(2,1)}\right\rangle_{\mathrm{c}}\left\langle q_{(1,1)} q_{(2,2)}\right\rangle_{\mathrm{c}}\right)\\
&\hphantom{\quad\ }+6\left\langle q_{(1,1)} q_{(2,1)} q_{(3,1)}\right\rangle_{\mathrm{c}}-18\left\langle q_{(1,1)} q_{(2,1)} q_{(3,2)}\right\rangle_{\mathrm{c}}+12\left\langle q_{(1,1)} q_{(2,1)} q_{(3,3)}\right\rangle_{\mathrm{c}}\\
&\hphantom{\quad\ }-6\left\langle q_{(1,1)} q_{(2,2)} q_{(3,1)}\right\rangle_{\mathrm{c}}+18\left\langle q_{(1,1)} q_{(2,2)} q_{(3,2)}\right\rangle_{\mathrm{c}}-12\left\langle q_{(1,1)} q_{(2,2)} q_{(3,3)}\right\rangle_{\mathrm{c}}\\
&\hphantom{\quad\ }-6\left\langle q_{(3,1)} q_{(3,2)}\right\rangle_{\mathrm{c}}+4\left\langle q_{(3,1)} q_{(3,3)}\right\rangle_{\mathrm{c}}-12\left\langle q_{(3,2)} q_{(3,3)}\right\rangle_{\mathrm{c}}
\end{split}\\
n\cdot\mathrm{Var}\left(\widehat{C}_4^{\mathrm{corr}}\right)&=\left\langle q_{(1,1)}^8\right\rangle_\mathrm{c}+16\left\langle q_{(1,1)}^6\right\rangle_\mathrm{c}\left\langle q_{(1,1)}^2\right\rangle_\mathrm{c}+48\left\langle q_{(1,1)}^5\right\rangle_\mathrm{c}\left\langle q_{(1,1)}^3\right\rangle_\mathrm{c}+34\left\langle q_{(1,1)}^4\right\rangle_\mathrm{c}^2\nonumber\\
&\hphantom{\quad\ }+72\left\langle q_{(1,1)}^4\right\rangle_\mathrm{c}\left\langle q_{(1,1)}^2\right\rangle_\mathrm{c}^2+144\left\langle q_{(1,1)}^3\right\rangle_\mathrm{c}^2\left\langle q_{(1,1)}^2\right\rangle_\mathrm{c}+24\left\langle q_{(1,1)}^2\right\rangle_\mathrm{c}^4\nonumber\\
&\hphantom{\quad\ }+36\left(\left\langle q_{(1,1)}^4q_{(2,1)}^2\right\rangle_\mathrm{c}+\left\langle q_{(1,1)}^4\right\rangle_\mathrm{c}\left\langle q_{(2,1)}^2\right\rangle_\mathrm{c}+4\left\langle q_{(1,1)}^3q_{(2,1)}\right\rangle_\mathrm{c}\left\langle q_{(1,1)}q_{(2,1)}\right\rangle_\mathrm{c}+4\left\langle q_{(1,1)}^2q_{(2,1)}^2\right\rangle_\mathrm{c}\left\langle q_{(1,1)}^2\right\rangle_\mathrm{c}\right.\nonumber\\
&\hphantom{\qquad\,}\left.+4\left\langle q_{(1,1)}^3\right\rangle_\mathrm{c}\left\langle q_{(1,1)}q_{(2,1)}^2\right\rangle_\mathrm{c}+5\left\langle q_{(1,1)}^2q_{(2,1)}\right\rangle_\mathrm{c}^2+2\left\langle q_{(1,1)}^2\right\rangle_\mathrm{c}^2\left\langle q_{(2,1)}^2\right\rangle_\mathrm{c}+4\left\langle q_{(1,1)}^2\right\rangle_\mathrm{c}\left\langle q_{(1,1)}q_{(2,1)}\right\rangle_\mathrm{c}^2\right)\nonumber\\
&\hphantom{\quad\ }+36\left(\left\langle q_{(1,1)}^4q_{(2,2)}^2\right\rangle_\mathrm{c}+\left\langle q_{(1,1)}^4\right\rangle_\mathrm{c}\left\langle q_{(2,2)}^2\right\rangle_\mathrm{c}+4\left\langle q_{(1,1)}^3q_{(2,2)}\right\rangle_\mathrm{c}\left\langle q_{(1,1)}q_{(2,2)}\right\rangle_\mathrm{c}+4\left\langle q_{(1,1)}^2q_{(2,2)}^2\right\rangle_\mathrm{c}\left\langle q_{(1,1)}^2\right\rangle_\mathrm{c}\right.\nonumber\\
&\hphantom{\qquad\,}\left.+4\left\langle q_{(1,1)}^3\right\rangle_\mathrm{c}\left\langle q_{(1,1)}q_{(2,2)}^2\right\rangle_\mathrm{c}+5\left\langle q_{(1,1)}^2q_{(2,2)}\right\rangle_\mathrm{c}^2+2\left\langle q_{(1,1)}^2\right\rangle_\mathrm{c}^2\left\langle q_{(2,2)}^2\right\rangle_\mathrm{c}+4\left\langle q_{(1,1)}^2\right\rangle_\mathrm{c}\left\langle q_{(1,1)}q_{(2,2)}\right\rangle_\mathrm{c}^2\right)\nonumber\\
&\hphantom{\quad\ }+16\left(\left\langle q_{(1,1)}^2q_{(3,1)}^2\right\rangle_\mathrm{c}+\left\langle q_{(1,1)}^2\right\rangle_\mathrm{c}\left\langle q_{(3,1)}^2\right\rangle_\mathrm{c}+\left\langle q_{(1,1)}q_{(3,1)}\right\rangle_\mathrm{c}^2\right)\nonumber\\
&\hphantom{\quad\ }+144\left(\left\langle q_{(1,1)}^2q_{(3,2)}^2\right\rangle_\mathrm{c}+\left\langle q_{(1,1)}^2\right\rangle_\mathrm{c}\left\langle q_{(3,2)}^2\right\rangle_\mathrm{c}+\left\langle q_{(1,1)}q_{(3,2)}\right\rangle_\mathrm{c}^2\right)\nonumber\\
&\hphantom{\quad\ }+64\left(\left\langle q_{(1,1)}^2q_{(3,3)}^2\right\rangle_\mathrm{c}+\left\langle q_{(1,1)}^2\right\rangle_\mathrm{c}\left\langle q_{(3,3)}^2\right\rangle_\mathrm{c}+\left\langle q_{(1,1)}q_{(3,3)}\right\rangle_\mathrm{c}^2\right)\nonumber\\
&\hphantom{\quad\ }+9\left(\left\langle q_{(2,1)}^4\right\rangle_\mathrm{c}+2\left\langle q_{(2,1)}^2\right\rangle_\mathrm{c}^2\right)+36\left(\left\langle q_{(2,1)}^2q_{(2,2)}^2\right\rangle_\mathrm{c}+\left\langle q_{(2,1)}^2\right\rangle_\mathrm{c}\left\langle q_{(2,2)}^2\right\rangle_\mathrm{c}+\left\langle q_{(2,1)}q_{(2,2)}\right\rangle_\mathrm{c}^2\right)\nonumber\\
&\hphantom{\quad\ }+9\left(\left\langle q_{(2,2)}^4\right\rangle_\mathrm{c}+2\left\langle q_{(2,2)}^2\right\rangle_\mathrm{c}^2\right)+\left\langle q_{(4,1)}^2\right\rangle_\mathrm{c}+49\left\langle q_{(4,2)}^2\right\rangle_\mathrm{c}+144\left\langle q_{(4,3)}^2\right\rangle_\mathrm{c}+36\left\langle q_{(4,4)}^2\right\rangle_\mathrm{c}\nonumber\\
&\hphantom{\quad\ }+12\left(\left\langle q_{(1,1)}^6q_{(2,1)}\right\rangle_\mathrm{c}+4\left\langle q_{(1,1)}^5\right\rangle_\mathrm{c}\left\langle q_{(1,1)}q_{(2,1)}\right\rangle_\mathrm{c}+8\left\langle q_{(1,1)}^4q_{(2,1)}\right\rangle_\mathrm{c}\left\langle q_{(1,1)}^2\right\rangle_\mathrm{c}+14\left\langle q_{(1,1)}^4\right\rangle_\mathrm{c}\left\langle q_{(1,1)}^2q_{(2,1)}\right\rangle_\mathrm{c}\right.\nonumber\\
&\hphantom{\qquad\,}\left.+16\left\langle q_{(1,1)}^3q_{(2,1)}\right\rangle_\mathrm{c}\left\langle q_{(1,1)}^3\right\rangle_\mathrm{c}+24\left\langle q_{(1,1)}^3\right\rangle_\mathrm{c}\left\langle q_{(1,1)}^2\right\rangle_\mathrm{c}\left\langle q_{(1,1)}q_{(2,1)}\right\rangle_\mathrm{c}+12\left\langle q_{(1,1)}^2q_{(2,1)}\right\rangle_\mathrm{c}\left\langle q_{(1,1)}^2\right\rangle_\mathrm{c}^2\right)\nonumber\\
&\hphantom{\quad\ }-12\left(\left\langle q_{(1,1)}^6q_{(2,2)}\right\rangle_\mathrm{c}+4\left\langle q_{(1,1)}^5\right\rangle_\mathrm{c}\left\langle q_{(1,1)}q_{(2,2)}\right\rangle_\mathrm{c}+8\left\langle q_{(1,1)}^4q_{(2,2)}\right\rangle_\mathrm{c}\left\langle q_{(1,1)}^2\right\rangle_\mathrm{c}+14\left\langle q_{(1,1)}^4\right\rangle_\mathrm{c}\left\langle q_{(1,1)}^2q_{(2,2)}\right\rangle_\mathrm{c}\right.\nonumber\\
&\hphantom{\qquad\,}\left.+16\left\langle q_{(1,1)}^3q_{(2,2)}\right\rangle_\mathrm{c}\left\langle q_{(1,1)}^3\right\rangle_\mathrm{c}+24\left\langle q_{(1,1)}^3\right\rangle_\mathrm{c}\left\langle q_{(1,1)}^2\right\rangle_\mathrm{c}\left\langle q_{(1,1)}q_{(2,2)}\right\rangle_\mathrm{c}+12\left\langle q_{(1,1)}^2q_{(2,2)}\right\rangle_\mathrm{c}\left\langle q_{(1,1)}^2\right\rangle_\mathrm{c}^2\right)\nonumber\\
&\hphantom{\quad\ }+8\left(\left\langle q_{(1,1)}^5q_{(3,1)}\right\rangle_\mathrm{c}+4\left\langle q_{(1,1)}^4\right\rangle_\mathrm{c}\left\langle q_{(1,1)}q_{(3,1)}\right\rangle_\mathrm{c}+4\left\langle q_{(1,1)}^3q_{(3,1)}\right\rangle_\mathrm{c}\left\langle q_{(1,1)}^2\right\rangle_\mathrm{c}+6\left\langle q_{(1,1)}^3\right\rangle_\mathrm{c}\left\langle q_{(1,1)}^2q_{(3,1)}\right\rangle_\mathrm{c}\right)\nonumber\\
&\hphantom{\quad\ }-24\left(\left\langle q_{(1,1)}^5q_{(3,2)}\right\rangle_\mathrm{c}+4\left\langle q_{(1,1)}^4\right\rangle_\mathrm{c}\left\langle q_{(1,1)}q_{(3,2)}\right\rangle_\mathrm{c}+4\left\langle q_{(1,1)}^3q_{(3,2)}\right\rangle_\mathrm{c}\left\langle q_{(1,1)}^2\right\rangle_\mathrm{c}+6\left\langle q_{(1,1)}^3\right\rangle_\mathrm{c}\left\langle q_{(1,1)}^2q_{(3,2)}\right\rangle_\mathrm{c}\right)\nonumber\\
&\hphantom{\quad\ }+16\left(\left\langle q_{(1,1)}^5q_{(3,3)}\right\rangle_\mathrm{c}+4\left\langle q_{(1,1)}^4\right\rangle_\mathrm{c}\left\langle q_{(1,1)}q_{(3,3)}\right\rangle_\mathrm{c}+4\left\langle q_{(1,1)}^3q_{(3,3)}\right\rangle_\mathrm{c}\left\langle q_{(1,1)}^2\right\rangle_\mathrm{c}+6\left\langle q_{(1,1)}^3\right\rangle_\mathrm{c}\left\langle q_{(1,1)}^2q_{(3,3)}\right\rangle_\mathrm{c}\right)\nonumber\\
&\hphantom{\quad\ }+6\left(\left\langle q_{(1,1)}^4q_{(2,1)}^2\right\rangle_\mathrm{c}+8\left\langle q_{(1,1)}^3q_{(2,1)}\right\rangle_\mathrm{c}\left\langle q_{(1,1)}q_{(2,1)}\right\rangle_\mathrm{c}+6\left\langle q_{(1,1)}^2q_{(2,1)}\right\rangle_\mathrm{c}^2\right)\nonumber\\
&\hphantom{\quad\ }-12\left(\left\langle q_{(1,1)}^4q_{(2,1)}q_{(2,2)}\right\rangle_\mathrm{c}+4\left\langle q_{(1,1)}^3q_{(2,1)}\right\rangle_\mathrm{c}\left\langle q_{(1,1)}q_{(2,2)}\right\rangle_\mathrm{c}\right.\nonumber\\
&\hphantom{\qquad\,}\left.+4\left\langle q_{(1,1)}^3q_{(2,2)}\right\rangle_\mathrm{c}\left\langle q_{(1,1)}q_{(2,1)}\right\rangle_\mathrm{c}+6\left\langle q_{(1,1)}^2q_{(2,1)}\right\rangle_\mathrm{c}\left\langle q_{(1,1)}^2q_{(2,2)}\right\rangle_\mathrm{c}\right)\nonumber\\
&\hphantom{\quad\ }+6\left(\left\langle q_{(1,1)}^4q_{(2,2)}^2\right\rangle_\mathrm{c}+8\left\langle q_{(1,1)}^3q_{(2,2)}\right\rangle_\mathrm{c}\left\langle q_{(1,1)}q_{(2,2)}\right\rangle_\mathrm{c}+6\left\langle q_{(1,1)}^2q_{(2,2)}\right\rangle_\mathrm{c}^2\right)\nonumber\\
&\hphantom{\quad\ }+2\left\langle q_{(1,1)}^4q_{(4,1)}\right\rangle_\mathrm{c}-14\left\langle q_{(1,1)}^4q_{(4,2)}\right\rangle_\mathrm{c}+24\left\langle q_{(1,1)}^4q_{(4,3)}\right\rangle_\mathrm{c}-12\left\langle q_{(1,1)}^4q_{(4,4)}\right\rangle_\mathrm{c}\nonumber\\
&\hphantom{\quad\ }-72\left(\left\langle q_{(1,1)}^4q_{(2,1)}q_{(2,2)}\right\rangle_\mathrm{c}+\left\langle q_{(1,1)}^4\right\rangle_\mathrm{c}\left\langle q_{(2,1)}q_{(2,2)}\right\rangle_\mathrm{c}+2\left\langle q_{(1,1)}^3q_{(2,1)}\right\rangle_\mathrm{c}\left\langle q_{(1,1)}q_{(2,2)}\right\rangle_\mathrm{c}\right.\nonumber\\
&\hphantom{\qquad\,}\left.+2\left\langle q_{(1,1)}^3q_{(2,2)}\right\rangle_\mathrm{c}\left\langle q_{(1,1)}q_{(2,1)}\right\rangle_\mathrm{c}+4\left\langle q_{(1,1)}^2q_{(2,1)}q_{(2,2)}\right\rangle_\mathrm{c}\left\langle q_{(1,1)}^2\right\rangle_\mathrm{c}+4\left\langle q_{(1,1)}^3\right\rangle_\mathrm{c}\left\langle q_{(1,1)}q_{(2,1)}q_{(2,2)}\right\rangle_\mathrm{c}\right.\nonumber\\
&\hphantom{\qquad\,}\left.+5\left\langle q_{(1,1)}^2q_{(2,1)}\right\rangle_\mathrm{c}\left\langle q_{(1,1)}^2q_{(2,2)}\right\rangle_\mathrm{c}+2\left\langle q_{(1,1)}^2\right\rangle_\mathrm{c}^2\left\langle q_{(2,1)}q_{(2,2)}\right\rangle_\mathrm{c}+4\left\langle q_{(1,1)}^2\right\rangle_\mathrm{c}\left\langle q_{(1,1)}q_{(2,1)}\right\rangle_\mathrm{c}\left\langle q_{(1,1)}q_{(2,2)}\right\rangle_\mathrm{c}\right)\nonumber\\
&\hphantom{\quad\ }+48\left(\left\langle q_{(1,1)}^3q_{(2,1)}q_{(3,1)}\right\rangle_\mathrm{c}+2\left\langle q_{(1,1)}^2\right\rangle_\mathrm{c}\left\langle q_{(1,1)}q_{(2,1)}q_{(3,1)}\right\rangle_\mathrm{c}+\left\langle q_{(1,1)}^2q_{(3,1)}\right\rangle_\mathrm{c}\left\langle q_{(1,1)}q_{(2,1)}\right\rangle_\mathrm{c}\right.\nonumber\\
&\hphantom{\qquad\,}\left.+2\left\langle q_{(1,1)}^2q_{(2,1)}\right\rangle_\mathrm{c}\left\langle q_{(1,1)}q_{(3,1)}\right\rangle_\mathrm{c}+\left\langle q_{(1,1)}^3\right\rangle_\mathrm{c}\left\langle q_{(2,1)}q_{(3,1)}\right\rangle_\mathrm{c}\right)\nonumber\\
&\hphantom{\quad\ }-144\left(\left\langle q_{(1,1)}^3q_{(2,1)}q_{(3,2)}\right\rangle_\mathrm{c}+2\left\langle q_{(1,1)}^2\right\rangle_\mathrm{c}\left\langle q_{(1,1)}q_{(2,1)}q_{(3,2)}\right\rangle_\mathrm{c}+\left\langle q_{(1,1)}^2q_{(3,2)}\right\rangle_\mathrm{c}\left\langle q_{(1,1)}q_{(2,1)}\right\rangle_\mathrm{c}\right.\nonumber\\
&\hphantom{\qquad\,}\left.+2\left\langle q_{(1,1)}^2q_{(2,1)}\right\rangle_\mathrm{c}\left\langle q_{(1,1)}q_{(3,2)}\right\rangle_\mathrm{c}+\left\langle q_{(1,1)}^3\right\rangle_\mathrm{c}\left\langle q_{(2,1)}q_{(3,2)}\right\rangle_\mathrm{c}\right)\nonumber\\
&\hphantom{\quad\ }+96\left(\left\langle q_{(1,1)}^3q_{(2,1)}q_{(3,3)}\right\rangle_\mathrm{c}+2\left\langle q_{(1,1)}^2\right\rangle_\mathrm{c}\left\langle q_{(1,1)}q_{(2,1)}q_{(3,3)}\right\rangle_\mathrm{c}+\left\langle q_{(1,1)}^2q_{(3,3)}\right\rangle_\mathrm{c}\left\langle q_{(1,1)}q_{(2,1)}\right\rangle_\mathrm{c}\right.\nonumber\\
&\hphantom{\qquad\,}\left.+2\left\langle q_{(1,1)}^2q_{(2,1)}\right\rangle_\mathrm{c}\left\langle q_{(1,1)}q_{(3,3)}\right\rangle_\mathrm{c}+\left\langle q_{(1,1)}^3\right\rangle_\mathrm{c}\left\langle q_{(2,1)}q_{(3,3)}\right\rangle_\mathrm{c}\right)\nonumber\\
&\hphantom{\quad\ }+36\left(\left\langle q_{(1,1)}^2q_{(2,1)}^3\right\rangle_\mathrm{c}+2\left\langle q_{(1,1)}^2q_{(2,1)}\right\rangle_\mathrm{c}\left\langle q_{(2,1)}^2\right\rangle_\mathrm{c}+4\left\langle q_{(1,1)}q_{(2,1)}^2\right\rangle_\mathrm{c}\left\langle q_{(1,1)}q_{(2,1)}\right\rangle_\mathrm{c}\right)\nonumber\\
&\hphantom{\quad\ }-72\left(\left\langle q_{(1,1)}^2q_{(2,1)}^2q_{(2,2)}\right\rangle_\mathrm{c}+\left\langle q_{(1,1)}^2q_{(2,1)}\right\rangle_\mathrm{c}\left\langle q_{(2,1)}q_{(2,2)}\right\rangle_\mathrm{c}+\left\langle q_{(1,1)}^2q_{(2,2)}\right\rangle_\mathrm{c}\left\langle q_{(2,1)}^2\right\rangle_\mathrm{c}\right.\nonumber\\
&\hphantom{\qquad\,}\left.+2\left\langle q_{(1,1)}q_{(2,1)}^2\right\rangle_\mathrm{c}\left\langle q_{(1,1)}q_{(2,2)}\right\rangle_\mathrm{c}+2\left\langle q_{(1,1)}q_{(2,1)}q_{(2,2)}\right\rangle_\mathrm{c}\left\langle q_{(1,1)}q_{(2,1)}\right\rangle_\mathrm{c}\right)\nonumber\\
&\hphantom{\quad\ }+36\left(\left\langle q_{(1,1)}^2q_{(2,1)}q_{(2,2)}^2\right\rangle_\mathrm{c}+2\left\langle q_{(1,1)}^2q_{(2,2)}\right\rangle_\mathrm{c}\left\langle q_{(2,1)}q_{(2,2)}\right\rangle_\mathrm{c}+4\left\langle q_{(1,1)}q_{(2,1)}q_{(2,2)}\right\rangle_\mathrm{c}\left\langle q_{(1,1)}q_{(2,2)}\right\rangle_\mathrm{c}\right)\nonumber\\
&\hphantom{\quad\ }+12\left\langle q_{(1,1)}^2q_{(2,1)}q_{(4,1)}\right\rangle_\mathrm{c}-84\left\langle q_{(1,1)}^2q_{(2,1)}q_{(4,2)}\right\rangle_\mathrm{c}+144\left\langle q_{(1,1)}^2q_{(2,1)}q_{(4,3)}\right\rangle_\mathrm{c}-72\left\langle q_{(1,1)}^2q_{(2,1)}q_{(4,4)}\right\rangle_\mathrm{c}\nonumber\\
&\hphantom{\quad\ }-48\left(\left\langle q_{(1,1)}^3q_{(2,2)}q_{(3,1)}\right\rangle_\mathrm{c}+2\left\langle q_{(1,1)}^2\right\rangle_\mathrm{c}\left\langle q_{(1,1)}q_{(2,2)}q_{(3,1)}\right\rangle_\mathrm{c}+\left\langle q_{(1,1)}^2q_{(3,1)}\right\rangle_\mathrm{c}\left\langle q_{(1,1)}q_{(2,2)}\right\rangle_\mathrm{c}\right.\nonumber\\
&\hphantom{\qquad\,}\left.+2\left\langle q_{(1,1)}^2q_{(2,2)}\right\rangle_\mathrm{c}\left\langle q_{(1,1)}q_{(3,1)}\right\rangle_\mathrm{c}+\left\langle q_{(1,1)}^3\right\rangle_\mathrm{c}\left\langle q_{(2,2)}q_{(3,1)}\right\rangle_\mathrm{c}\right)\nonumber\\
&\hphantom{\quad\ }+144\left(\left\langle q_{(1,1)}^3q_{(2,2)}q_{(3,2)}\right\rangle_\mathrm{c}+2\left\langle q_{(1,1)}^2\right\rangle_\mathrm{c}\left\langle q_{(1,1)}q_{(2,2)}q_{(3,2)}\right\rangle_\mathrm{c}+\left\langle q_{(1,1)}^2q_{(3,2)}\right\rangle_\mathrm{c}\left\langle q_{(1,1)}q_{(2,2)}\right\rangle_\mathrm{c}\right.\nonumber\\
&\hphantom{\qquad\,}\left.+2\left\langle q_{(1,1)}^2q_{(2,2)}\right\rangle_\mathrm{c}\left\langle q_{(1,1)}q_{(3,2)}\right\rangle_\mathrm{c}+\left\langle q_{(1,1)}^3\right\rangle_\mathrm{c}\left\langle q_{(2,2)}q_{(3,2)}\right\rangle_\mathrm{c}\right)\nonumber\\
&\hphantom{\quad\ }-96\left(\left\langle q_{(1,1)}^3q_{(2,2)}q_{(3,3)}\right\rangle_\mathrm{c}+2\left\langle q_{(1,1)}^2\right\rangle_\mathrm{c}\left\langle q_{(1,1)}q_{(2,2)}q_{(3,3)}\right\rangle_\mathrm{c}+\left\langle q_{(1,1)}^2q_{(3,3)}\right\rangle_\mathrm{c}\left\langle q_{(1,1)}q_{(2,2)}\right\rangle_\mathrm{c}\right.\nonumber\\
&\hphantom{\qquad\,}\left.+2\left\langle q_{(1,1)}^2q_{(2,2)}\right\rangle_\mathrm{c}\left\langle q_{(1,1)}q_{(3,3)}\right\rangle_\mathrm{c}+\left\langle q_{(1,1)}^3\right\rangle_\mathrm{c}\left\langle q_{(2,2)}q_{(3,3)}\right\rangle_\mathrm{c}\right)\nonumber\\
&\hphantom{\quad\ }-36\left(\left\langle q_{(1,1)}^2q_{(2,1)}^2q_{(2,2)}\right\rangle_\mathrm{c}+2\left\langle q_{(1,1)}^2q_{(2,1)}\right\rangle_\mathrm{c}\left\langle q_{(2,1)}q_{(2,2)}\right\rangle_\mathrm{c}+4\left\langle q_{(1,1)}q_{(2,1)}q_{(2,2)}\right\rangle_\mathrm{c}\left\langle q_{(1,1)}q_{(2,1)}\right\rangle_\mathrm{c}\right)\nonumber\\
&\hphantom{\quad\ }+72\left(\left\langle q_{(1,1)}^2q_{(2,1)}q_{(2,2)}^2\right\rangle_\mathrm{c}+\left\langle q_{(1,1)}^2q_{(2,1)}\right\rangle_\mathrm{c}\left\langle q_{(2,2)}^2\right\rangle_\mathrm{c}+\left\langle q_{(1,1)}^2q_{(2,2)}\right\rangle_\mathrm{c}\left\langle q_{(2,1)}q_{(2,2)}\right\rangle_\mathrm{c}\right.\nonumber\\
&\hphantom{\qquad\,}\left.+2\left\langle q_{(1,1)}q_{(2,1)}q_{(2,2)}\right\rangle_\mathrm{c}\left\langle q_{(1,1)}q_{(2,2)}\right\rangle_\mathrm{c}+2\left\langle q_{(1,1)}q_{(2,1)}\right\rangle_\mathrm{c}\left\langle q_{(1,1)}q_{(2,2)}^2\right\rangle_\mathrm{c}\right)\nonumber\\
&\hphantom{\quad\ }-36\left(\left\langle q_{(1,1)}^2q_{(2,2)}^3\right\rangle_\mathrm{c}+2\left\langle q_{(1,1)}^2q_{(2,2)}\right\rangle_\mathrm{c}\left\langle q_{(2,2)}^2\right\rangle_\mathrm{c}+4\left\langle q_{(1,1)}q_{(2,2)}^2\right\rangle_\mathrm{c}\left\langle q_{(1,1)}q_{(2,2)}\right\rangle_\mathrm{c}\right)\nonumber\\
&\hphantom{\quad\ }-12\left\langle q_{(1,1)}^2q_{(2,2)}q_{(4,1)}\right\rangle_\mathrm{c}+84\left\langle q_{(1,1)}^2q_{(2,2)}q_{(4,2)}\right\rangle_\mathrm{c}-144\left\langle q_{(1,1)}^2q_{(2,2)}q_{(4,3)}\right\rangle_\mathrm{c}+72\left\langle q_{(1,1)}^2q_{(2,2)}q_{(4,4)}\right\rangle_\mathrm{c}\nonumber\\
&\hphantom{\quad\ }-96\left(\left\langle q_{(1,1)}^2q_{(3,1)}q_{(3,2)}\right\rangle_\mathrm{c}+\left\langle q_{(1,1)}^2\right\rangle_\mathrm{c}\left\langle q_{(3,1)}q_{(3,2)}\right\rangle_\mathrm{c}+\left\langle q_{(1,1)}q_{(3,1)}\right\rangle_\mathrm{c}\left\langle q_{(1,1)}q_{(3,2)}\right\rangle_\mathrm{c}\right)\nonumber\\
&\hphantom{\quad\ }+64\left(\left\langle q_{(1,1)}^2q_{(3,1)}q_{(3,3)}\right\rangle_\mathrm{c}+\left\langle q_{(1,1)}^2\right\rangle_\mathrm{c}\left\langle q_{(3,1)}q_{(3,3)}\right\rangle_\mathrm{c}+\left\langle q_{(1,1)}q_{(3,1)}\right\rangle_\mathrm{c}\left\langle q_{(1,1)}q_{(3,3)}\right\rangle_\mathrm{c}\right)\nonumber\\
&\hphantom{\quad\ }+24\left(\left\langle q_{(1,1)}q_{(2,1)}^2q_{(3,1)}\right\rangle_\mathrm{c}+2\left\langle q_{(1,1)}q_{(2,1)}\right\rangle_\mathrm{c}\left\langle q_{(2,1)}q_{(3,1)}\right\rangle_\mathrm{c}\right)\nonumber\\
&\hphantom{\quad\ }-48\left(\left\langle q_{(1,1)}q_{(2,1)}q_{(2,2)}q_{(3,1)}\right\rangle_\mathrm{c}+\left\langle q_{(1,1)}q_{(2,1)}\right\rangle_\mathrm{c}\left\langle q_{(2,2)}q_{(3,1)}\right\rangle_\mathrm{c}+\left\langle q_{(1,1)}q_{(2,2)}\right\rangle_\mathrm{c}\left\langle q_{(2,1)}q_{(3,1)}\right\rangle_\mathrm{c}\right)\nonumber\\
&\hphantom{\quad\ }+24\left(\left\langle q_{(1,1)}q_{(2,2)}^2q_{(3,1)}\right\rangle_\mathrm{c}+2\left\langle q_{(1,1)}q_{(2,2)}\right\rangle_\mathrm{c}\left\langle q_{(2,2)}q_{(3,1)}\right\rangle_\mathrm{c}\right)\nonumber\\
&\hphantom{\quad\ }+8\left\langle q_{(1,1)}q_{(3,1)}q_{(4,1)}\right\rangle_\mathrm{c}-56\left\langle q_{(1,1)}q_{(3,1)}q_{(4,2)}\right\rangle_\mathrm{c}+96\left\langle q_{(1,1)}q_{(3,1)}q_{(4,3)}\right\rangle_\mathrm{c}-48\left\langle q_{(1,1)}q_{(3,1)}q_{(4,4)}\right\rangle_\mathrm{c}\nonumber\\
&\hphantom{\quad\ }-192\left(\left\langle q_{(1,1)}^2q_{(3,2)}q_{(3,3)}\right\rangle_\mathrm{c}+\left\langle q_{(1,1)}^2\right\rangle_\mathrm{c}\left\langle q_{(3,2)}q_{(3,3)}\right\rangle_\mathrm{c}+\left\langle q_{(1,1)}q_{(3,2)}\right\rangle_\mathrm{c}\left\langle q_{(1,1)}q_{(3,3)}\right\rangle_\mathrm{c}\right)\nonumber\\
&\hphantom{\quad\ }-72\left(\left\langle q_{(1,1)}q_{(2,1)}^2q_{(3,2)}\right\rangle_\mathrm{c}+2\left\langle q_{(1,1)}q_{(2,1)}\right\rangle_\mathrm{c}\left\langle q_{(2,1)}q_{(3,2)}\right\rangle_\mathrm{c}\right)\nonumber\\
&\hphantom{\quad\ }+144\left(\left\langle q_{(1,1)}q_{(2,1)}q_{(2,2)}q_{(3,2)}\right\rangle_\mathrm{c}+\left\langle q_{(1,1)}q_{(2,1)}\right\rangle_\mathrm{c}\left\langle q_{(2,2)}q_{(3,2)}\right\rangle_\mathrm{c}+\left\langle q_{(1,1)}q_{(2,2)}\right\rangle_\mathrm{c}\left\langle q_{(2,1)}q_{(3,2)}\right\rangle_\mathrm{c}\right)\nonumber\\
&\hphantom{\quad\ }-72\left(\left\langle q_{(1,1)}q_{(2,2)}^2q_{(3,2)}\right\rangle_\mathrm{c}+2\left\langle q_{(1,1)}q_{(2,2)}\right\rangle_\mathrm{c}\left\langle q_{(2,2)}q_{(3,2)}\right\rangle_\mathrm{c}\right)\nonumber\\
&\hphantom{\quad\ }-24\left\langle q_{(1,1)}q_{(3,2)}q_{(4,1)}\right\rangle_\mathrm{c}+168\left\langle q_{(1,1)}q_{(3,2)}q_{(4,2)}\right\rangle_\mathrm{c}-288\left\langle q_{(1,1)}q_{(3,2)}q_{(4,3)}\right\rangle_\mathrm{c}+144\left\langle q_{(1,1)}q_{(3,2)}q_{(4,4)}\right\rangle_\mathrm{c}\nonumber\\
&\hphantom{\quad\ }+48\left(\left\langle q_{(1,1)}q_{(2,1)}^2q_{(3,3)}\right\rangle_\mathrm{c}+2\left\langle q_{(1,1)}q_{(2,1)}\right\rangle_\mathrm{c}\left\langle q_{(2,1)}q_{(3,3)}\right\rangle_\mathrm{c}\right)\nonumber\\
&\hphantom{\quad\ }-96\left(\left\langle q_{(1,1)}q_{(2,1)}q_{(2,2)}q_{(3,3)}\right\rangle_\mathrm{c}+\left\langle q_{(1,1)}q_{(2,1)}\right\rangle_\mathrm{c}\left\langle q_{(2,2)}q_{(3,3)}\right\rangle_\mathrm{c}+\left\langle q_{(1,1)}q_{(2,2)}\right\rangle_\mathrm{c}\left\langle q_{(2,1)}q_{(3,3)}\right\rangle_\mathrm{c}\right)\nonumber\\
&\hphantom{\quad\ }+48\left(\left\langle q_{(1,1)}q_{(2,2)}^2q_{(3,3)}\right\rangle_\mathrm{c}+2\left\langle q_{(1,1)}q_{(2,2)}\right\rangle_\mathrm{c}\left\langle q_{(2,2)}q_{(3,3)}\right\rangle_\mathrm{c}\right)\nonumber\\
&\hphantom{\quad\ }+16\left\langle q_{(1,1)}q_{(3,3)}q_{(4,1)}\right\rangle_\mathrm{c}-112\left\langle q_{(1,1)}q_{(3,3)}q_{(4,2)}\right\rangle_\mathrm{c}+192\left\langle q_{(1,1)}q_{(3,3)}q_{(4,3)}\right\rangle_\mathrm{c}-96\left\langle q_{(1,1)}q_{(3,3)}q_{(4,4)}\right\rangle_\mathrm{c}\nonumber\\
&\hphantom{\quad\ }-36\left(\left\langle q_{(2,1)}^3q_{(2,2)}\right\rangle_\mathrm{c}+2\left\langle q_{(2,1)}^2\right\rangle_\mathrm{c}\left\langle q_{(2,1)}q_{(2,2)}\right\rangle_\mathrm{c}\right)+18\left(\left\langle q_{(2,1)}^2q_{(2,2)}^2\right\rangle_\mathrm{c}+2\left\langle q_{(2,1)}q_{(2,2)}\right\rangle_\mathrm{c}^2\right)\nonumber\\
&\hphantom{\quad\ }+6\left\langle q_{(2,1)}^2q_{(4,1)}\right\rangle_\mathrm{c}-42\left\langle q_{(2,1)}^2q_{(4,2)}\right\rangle_\mathrm{c}+72\left\langle q_{(2,1)}^2q_{(4,3)}\right\rangle_\mathrm{c}-36\left\langle q_{(2,1)}^2q_{(4,4)}\right\rangle_\mathrm{c}\nonumber\\
&\hphantom{\quad\ }-36\left(\left\langle q_{(2,1)}q_{(2,2)}^3\right\rangle_\mathrm{c}+2\left\langle q_{(2,1)}q_{(2,2)}\right\rangle_\mathrm{c}\left\langle q_{(2,2)}^2\right\rangle_\mathrm{c}\right)\nonumber\\
&\hphantom{\quad\ }-12\left\langle q_{(2,1)}q_{(2,2)}q_{(4,1)}\right\rangle_\mathrm{c}+84\left\langle q_{(2,1)}q_{(2,2)}q_{(4,2)}\right\rangle_\mathrm{c}-144\left\langle q_{(2,1)}q_{(2,2)}q_{(4,3)}\right\rangle_\mathrm{c}+72\left\langle q_{(2,1)}q_{(2,2)}q_{(4,4)}\right\rangle_\mathrm{c}\nonumber\\
&\hphantom{\quad\ }+6\left\langle q_{(2,2)}^2q_{(4,1)}\right\rangle_\mathrm{c}-42\left\langle q_{(2,2)}^2q_{(4,2)}\right\rangle_\mathrm{c}+72\left\langle q_{(2,2)}^2q_{(4,3)}\right\rangle_\mathrm{c}-36\left\langle q_{(2,2)}^2q_{(4,4)}\right\rangle_\mathrm{c}\nonumber\\
&\hphantom{\quad\ }-14\left\langle q_{(4,1)}q_{(4,2)}\right\rangle_\mathrm{c}+24\left\langle q_{(4,1)}q_{(4,3)}\right\rangle_\mathrm{c}-12\left\langle q_{(4,1)}q_{(4,4)}\right\rangle_\mathrm{c}\nonumber\\
&\hphantom{\quad\ }-168\left\langle q_{(4,2)}q_{(4,3)}\right\rangle_\mathrm{c}+84\left\langle q_{(4,2)}q_{(4,4)}\right\rangle_\mathrm{c}-144\left\langle q_{(4,3)}q_{(4,4)}\right\rangle_\mathrm{c}
\label{eqe4}
\end{align}

Some covariances between estimated efficiency-corrected cumulants in the multivariate case are shown below.
\begin{align}
\label{eqer3}
\begin{split}
n\cdot\mathrm{Cov}\left(\widehat{C}_3^{\mathrm{corr}},\widehat{C}_2^{\mathrm{corr}}\right)&=\left\langle q_{(1,1)}^5\right\rangle_{\mathrm{c}}+6\left\langle q_{(1,1)}^3\right\rangle_{\mathrm{c}}\left\langle q_{(1,1)}^2\right\rangle_{\mathrm{c}}+3\left(\left\langle q_{(1,1)}^3q_{(2,1)}\right\rangle_{\mathrm{c}}+2\left\langle q_{(1,1)}^2\right\rangle_{\mathrm{c}}\left\langle q_{(1,1)}q_{(2,1)}\right\rangle_{\mathrm{c}}\right)\\
&\hphantom{\quad\ }-3\left(\left\langle q_{(1,1)}^3q_{(2,2)}\right\rangle_{\mathrm{c}}+2\left\langle q_{(1,1)}^2\right\rangle_{\mathrm{c}}\left\langle q_{(1,1)}q_{(2,2)}\right\rangle_{\mathrm{c}}\right)+\left\langle q_{(1,1)}^2q_{(3,1)}\right\rangle_{\mathrm{c}}-3\left\langle q_{(1,1)}^2q_{(3,2)}\right\rangle_{\mathrm{c}}\\
&\hphantom{\quad\ }+2\left\langle q_{(1,1)}^2q_{(3,3)}\right\rangle_{\mathrm{c}}+\left\langle q_{(1,1)}^3q_{(2,1)}\right\rangle_{\mathrm{c}}+3\left\langle q_{(1,1)}q_{(2,1)}^2\right\rangle_{\mathrm{c}}-3\left\langle q_{(1,1)}q_{(2,1)}q_{(2,2)}\right\rangle_{\mathrm{c}}\\
&\hphantom{\quad\ }+\left\langle q_{(2,1)}q_{(3,1)}\right\rangle_{\mathrm{c}}-3\left\langle q_{(2,1)}q_{(3,2)}\right\rangle_{\mathrm{c}}+2\left\langle q_{(2,1)}q_{(3,3)}\right\rangle_{\mathrm{c}}-\left\langle q_{(1,1)}^3q_{(2,2)}\right\rangle_{\mathrm{c}}-3\left\langle q_{(1,1)}q_{(2,1)}q_{(2,2)}\right\rangle_{\mathrm{c}}
\end{split}\nonumber\\
&\hphantom{\quad\ }+3\left\langle q_{(1,1)}q_{(2,2)}^2\right\rangle_{\mathrm{c}}-\left\langle q_{(2,2)}q_{(3,1)}\right\rangle_{\mathrm{c}}+3\left\langle q_{(2,2)}q_{(3,2)}\right\rangle_{\mathrm{c}}-2\left\langle q_{(2,2)}q_{(3,3)}\right\rangle_{\mathrm{c}}\\
\label{eqer4}
n\cdot\mathrm{Cov}\left(\widehat{C}_4^{\mathrm{corr}},\widehat{C}_2^{\mathrm{corr}}\right)&=\left\langle q_{(1,1)}^6\right\rangle_{\mathrm{c}}+8\left\langle q_{(1,1)}^4\right\rangle_{\mathrm{c}}\left\langle q_{(1,1)}^2\right\rangle_{\mathrm{c}}+6\left\langle q_{(1,1)}^3\right\rangle_{\mathrm{c}}^2\nonumber\\
&\hphantom{\quad\ }+6\left(\left\langle q_{(1,1)}^4q_{(2,1)}\right\rangle_{\mathrm{c}}+4\left\langle q_{(1,1)}^2q_{(2,1)}\right\rangle_{\mathrm{c}}\left\langle q_{(1,1)}^2\right\rangle_{\mathrm{c}}+2\left\langle q_{(1,1)}^3\right\rangle_{\mathrm{c}}\left\langle q_{(1,1)}q_{(2,1)}\right\rangle_{\mathrm{c}}\right)\nonumber\\
&\hphantom{\quad\ }-6\left(\left\langle q_{(1,1)}^4q_{(2,2)}\right\rangle_{\mathrm{c}}+4\left\langle q_{(1,1)}^2q_{(2,2)}\right\rangle_{\mathrm{c}}\left\langle q_{(1,1)}^2\right\rangle_{\mathrm{c}}+2\left\langle q_{(1,1)}^3\right\rangle_{\mathrm{c}}\left\langle q_{(1,1)}q_{(2,2)}\right\rangle_{\mathrm{c}}\right)\nonumber\\
&\hphantom{\quad\ }+4\left(\left\langle q_{(1,1)}^3q_{(3,1)}\right\rangle_{\mathrm{c}}+2\left\langle q_{(1,1)}^2\right\rangle_{\mathrm{c}}\left\langle q_{(1,1)}q_{(3,1)}\right\rangle_{\mathrm{c}}\right)\nonumber\\
&\hphantom{\quad\ }-12\left(\left\langle q_{(1,1)}^3q_{(3,2)}\right\rangle_{\mathrm{c}}+2\left\langle q_{(1,1)}^2\right\rangle_{\mathrm{c}}\left\langle q_{(1,1)}q_{(3,2)}\right\rangle_{\mathrm{c}}\right)\nonumber\\
&\hphantom{\quad\ }+8\left(\left\langle q_{(1,1)}^3q_{(3,3)}\right\rangle_{\mathrm{c}}+2\left\langle q_{(1,1)}^2\right\rangle_{\mathrm{c}}\left\langle q_{(1,1)}q_{(3,3)}\right\rangle_{\mathrm{c}}\right)\nonumber\\
&\hphantom{\quad\ }+3\left(\left\langle q_{(1,1)}^2q_{(2,1)}^2\right\rangle_{\mathrm{c}}+2\left\langle q_{(1,1)}q_{(2,1)}\right\rangle_{\mathrm{c}}^2\right)\nonumber\\
&\hphantom{\quad\ }-6\left(\left\langle q_{(1,1)}^2q_{(2,1)}q_{(2,2)}\right\rangle_{\mathrm{c}}+2\left\langle q_{(1,1)}q_{(2,1)}\right\rangle_{\mathrm{c}}\left\langle q_{(1,1)}q_{(2,2)}\right\rangle_{\mathrm{c}}\right)\nonumber\\
&\hphantom{\quad\ }+3\left(\left\langle q_{(1,1)}^2q_{(2,2)}^2\right\rangle_{\mathrm{c}}+2\left\langle q_{(1,1)}q_{(2,2)}\right\rangle_{\mathrm{c}}^2\right)\nonumber\\
&\hphantom{\quad\ }+\left\langle q_{(1,1)}^2q_{(4,1)}\right\rangle_{\mathrm{c}}-7\left\langle q_{(1,1)}^2q_{(4,2)}\right\rangle_{\mathrm{c}}+12\left\langle q_{(1,1)}^2q_{(4,3)}\right\rangle_{\mathrm{c}}-6\left\langle q_{(1,1)}^2q_{(4,4)}\right\rangle_{\mathrm{c}}\nonumber\\
&\hphantom{\quad\ }+\left\langle q_{(1,1)}^4q_{(2,1)}\right\rangle_{\mathrm{c}}+6\left\langle q_{(1,1)}^2q_{(2,1)}^2\right\rangle_{\mathrm{c}}-6\left\langle q_{(1,1)}^2q_{(2,1)}q_{(2,2)}\right\rangle_{\mathrm{c}}+4\left\langle q_{(1,1)}q_{(2,1)}q_{(3,1)}\right\rangle_{\mathrm{c}}\nonumber\\
&\hphantom{\quad\ }-12\left\langle q_{(1,1)}q_{(2,1)}q_{(3,2)}\right\rangle_{\mathrm{c}}+8\left\langle q_{(1,1)}q_{(2,1)}q_{(3,3)}\right\rangle_{\mathrm{c}}+3\left\langle q_{(2,1)}^3\right\rangle_{\mathrm{c}}-6\left\langle q_{(2,1)}^2q_{(2,2)}\right\rangle_{\mathrm{c}}\nonumber\\
&\hphantom{\quad\ }+3\left\langle q_{(2,1)}q_{(2,2)}^2\right\rangle_{\mathrm{c}}+\left\langle q_{(2,1)}q_{(4,1)}\right\rangle_{\mathrm{c}}-7\left\langle q_{(2,1)}q_{(4,2)}\right\rangle_{\mathrm{c}}+12\left\langle q_{(2,1)}q_{(4,3)}\right\rangle_{\mathrm{c}}-6\left\langle q_{(2,1)}q_{(4,4)}\right\rangle_{\mathrm{c}}\nonumber\\
&\hphantom{\quad\ }-\left\langle q_{(1,1)}^4q_{(2,2)}\right\rangle_{\mathrm{c}}-6\left\langle q_{(1,1)}^2q_{(2,1)}q_{(2,2)}\right\rangle_{\mathrm{c}}+6\left\langle q_{(1,1)}^2q_{(2,2)}^2\right\rangle_{\mathrm{c}}-4\left\langle q_{(1,1)}q_{(2,2)}q_{(3,1)}\right\rangle_{\mathrm{c}}\nonumber\\
&\hphantom{\quad\ }+12\left\langle q_{(1,1)}q_{(2,2)}q_{(3,2)}\right\rangle_{\mathrm{c}}-8\left\langle q_{(1,1)}q_{(2,2)}q_{(3,3)}\right\rangle_{\mathrm{c}}-3\left\langle q_{(2,1)}^2q_{(2,2)}\right\rangle_{\mathrm{c}}+6\left\langle q_{(2,1)}q_{(2,2)}^2\right\rangle_{\mathrm{c}}\nonumber\\
&\hphantom{\quad\ }-3\left\langle q_{(2,2)}^3\right\rangle_{\mathrm{c}}-\left\langle q_{(2,2)}q_{(4,1)}\right\rangle_{\mathrm{c}}+7\left\langle q_{(2,2)}q_{(4,2)}\right\rangle_{\mathrm{c}}-12\left\langle q_{(2,2)}q_{(4,3)}\right\rangle_{\mathrm{c}}+6\left\langle q_{(2,2)}q_{(4,4)}\right\rangle_{\mathrm{c}}
\end{align}
\end{widetext}

\bibliographystyle{apsrev4-2}
\bibliography{references.bib}

\end{document}